\documentstyle[epsf,12pt]{article}
\newcommand{\bold}[1]{\mbox{\boldmath ${#1}$}}
\newcommand{\fslash}[1]{\mbox{$\!\not\!#1$}}

\newcommand{\be}{\begin{equation}}
\newcommand{\ee}{\end{equation}}

\newcommand{\kint}[1]{\int \! \!{d^4 #1 \over {(2\pi)^4}}}

\begin{document}
\baselineskip 4 ex
\title{The Stability of Nuclear Matter in the Nambu-Jona-Lasinio Model 
\thanks{
Preprint number: ADP-01-09/T444.  
E-mail addresses: bentz@keyaki.cc.u-tokai.ac.jp, athomas@physics.adelaide.edu.au}
\author{
W. Bentz \\
Dept. of Physics, School of Science, \\
Tokai University \\
1117 Kita-Kaname, Hiratsuka 259-1207, Japan \\
{ }\\
and \\
{ }\\
A. W. Thomas \\
Special Research Center for the Subatomic Structure of Matter \\
and \\
Department of Physics and Mathematical Physics, \\
Adelaide University \\
Adelaide, SA 5005, Australia}
}

\date{}
\maketitle
\newpage
\begin{abstract}
Using the Nambu-Jona-Lasinio model to describe the nucleon as a quark-diquark
state, we discuss the stability of nuclear matter in a hybrid model for the
ground state at finite nucleon density. It is shown that a simple extension of the
model to simulate the effects of confinement leads to a scalar polarizability of the nucleon.
This, in
turn, leads to a less attractive effective interaction between the nucleons,
helping to achieve saturation of the nuclear matter ground state.
It is also pointed out that that the same effect
naturally leads to a suppression of ``Z-graph'' contributions with increasing 
scalar potential.
\\

\noindent
{\small PACS: 12.39 Fe (chiral Lagrangians), 12.39 Ki (relativistic quark model), 21.65+f 
(nuclear matter), 24.85+p (quarks, gluons and QCD in nuclei and nuclear processes).\\
Keywords: quark-diquark structure of the nucleon, nuclear matter stability,
          Nambu-Jona-Lasinio model for quarks.}
\end{abstract}

\section{Introduction}
\setcounter{equation}{0}
Since the first applications of models based on the linear realization of
chiral symmetry to the description of nuclear many-body systems systems \cite{LEE}, 
the problem of the
saturation of the nuclear matter ground state in $\sigma$ \cite{SIGMA} or Nambu-Jona-Lasinio \cite{NJL}
(NJL) type models has been extensively discussed \cite{UNP}-\cite{BUB}. The common feature of these
models is a vacuum effective potential, which has a ``Mexican hat'' shape as a function
of the classical scalar field. The second derivative (curvature) of 
this effective potential decreases as one moves away from the minimum (vacuum state) towards smaller
scalar fields. This decrease of the curvature implies an attractive contribution to
the effective $\sigma$ meson 
mass at finite density as a result of vacuum fluctuation effects, and leads to an increase of the 
attraction between nucleons associated with $\sigma$ meson exchange. These attractive vacuum
fluctuation effects, which appear in the form of tadpole type diagrams in
$\sigma$-type models, do not
appear in non-chiral models like the phenomenologically successful Walecka model \cite{WAL}. 
They lead to an instability of the
nuclear matter ground state in the mean field (Hartree) approximation. 
Besides the possibilities of including higher order interaction terms in the Lagrangian \cite{BOG,KOCH},
or fluctuation terms arising from the Dirac sea of the nucleons \cite{WAL,LBA}, no practical methods towards
a solution of this problem have been found so far\footnote{Recently it has been argued \cite{BUB}
that the stable abnormal state of quark matter, which can be described in the NJL model,
should be interpreted as a droplet of massless quarks surrounded by the vacuum, i.e., as the
nucleon. While
this viewpoint is interesting, it seems very difficult to proceed to the description of
nuclear matter along these lines.}. Since the concept of the spontaneous breaking of the
chiral symmetry is exhibited most clearly and transparently in models based on the
linear realization of the symmetry, some hints towards a solution of this long standing
problem of matter stability would be highly welcome.

In recent years it has been shown that the Faddeev approach to the nucleon in the
NJL model is a very powerful method to achieve a covariant description of the
nucleon in terms of quarks and composite diquarks \cite{FAD1,FAD2,GROUND}. 
In particular, this method seems
to work very well for the description of the form factors \cite{FORM} and structure functions 
\cite{MIN} of a free nucleon. 
It would be interesting to extend this successful description
of the free nucleon to the investigation of medium modifications of nucleon properties, which 
is currently a very active field both experimentally and theoretically - an exciting example
are the recent studies of nucleon form factors in $(\vec{e},e'\vec{p})$ reactions \cite{FORM1}. For
this purpose, however, one needs the equation of state, and one first has to
solve the problem of the stability of nuclear matter.
A natural question in this respect is of course whether or not the quark substructure of 
the nucleon plays some role in producing saturation. Important hints
that this might be so come from the successful quark meson coupling model of Guichon 
and collaborators \cite{GUI}, where the saturation
arises from a relativistic effect associated with the quark motion inside the nucleon. 

Concerning the energy of a nucleon in a
self consistent scalar mean field, it has been pointed out \cite{ZGR} that there exists 
an important 
difference between an elementary and a composite nucleon. To second order in the scalar field, 
there is a repulsive contribution proportional to the square of the nucleon momentum in both cases. 
For the elementary nucleon case, this is the famous ``Z-graph'' contribution, but the presence of 
such a term is actually model independent as has been emphasized in a series of
papers \cite{ZGR,WALL}. It is very important for the description of nucleon-nucleus scattering and
for the description of saturation in the Walecka model \cite{BROWN}. It is also present in
chiral models, but at normal densities its contribution to the effective potential for nuclear matter 
is not large enough to stabilize the system against the attractive vacuum fluctuation
effects discussed above. Rather, the ``Z-graph'' contribution favors the formation of a stable 
abnormal state since it
increases strongly with increasing scalar potential (decreasing fermion mass).

For the case of a composite nucleon, however, there is in general an additional 
effective, contact-like, 
$\sigma^2 N^2$ coupling term, corresponding to the
``scalar polarizability'' of the nucleon. If this term is repulsive, it might play a role to stabilize 
the system. To investigate this
possibility in a simple NJL model calculation is the main motivation for our present work.
We demonstrate that, as long as there are unphysical thresholds in the NJL model for the decay of the
quark-diquark bound state, this
scalar polarizability is either too small or even of the wrong sign (attractive).
However, a sufficiently strong repulsive contribution can arise if confinement effects are
incorporated such as to avoid the unphysical thresholds. We show that such a term can lead to
a substantial reduction of the attraction arising from composite $\sigma$ meson exchange between two nucleons 
and therefore to saturation of the nuclear matter equation of state. We also
demonstrate that the same mechanism naturally leads to a suppression of the ``Z-graph''
contribution with increasing scalar potential (decreasing fermion mass).    

The rest of this paper is organized as follows: In sect. 2 we discuss a hybrid model for the
nuclear matter ground state based on the quark-diquark picture of a single nucleon. 
Several expansions in powers of the density are given, and 
the effective two-nucleon interaction (Landau-Migdal interaction) is discussed. 
The numerical results in sect. 3 show that, as long as the
model contains unphysical thresholds, it leads to essentially the same situation
as for elementary nucleons; i.e., it cannot describe saturating nuclear matter.
By using a simple method, based on an infrared cut-off, to avoid the unphysical thresholds
\cite{INFR}, 
it will be shown that the scalar polarizability of the nucleon discussed above does indeed work 
towards the stabilization of the system. However, for technical
reasons, in this work we will limit ourselves to the ``static approximation'' \cite{STAT1,STAT2}
to the full Faddeev equation
when describing the nucleon as a quark-diquark state. Since this approximation becomes worse for a nucleon in the medium
there remains an ambiguity which should be resolved in a future numerical analysis. 
Nevertheless, our results indicate that a more complete calculation, describing the nucleon in the scalar field by the
full Faddeev equation, will very probably lead to a saturating nuclear matter ground state. 
Finally, in sect. 4 we present discussions and conclusions.
      
\section{Nuclear matter in the NJL model}
\setcounter{equation}{0}
The NJL model is characterized by a chirally symmetric 4-fermi interaction between quarks.
Any such 4-fermi interaction can be Fierz symmetrized \cite{FAD1} and rewritten identically
as a chirally symmetric linear combination of the form $\sum_i G_i \left({\overline \psi}\Gamma_i 
\psi\right)^2$, where $\psi$ is the flavor SU(2) quark field, $\Gamma_i$ are matrices in Dirac, 
flavor and color space, and the coupling constants, $G_i$, are functions of the coupling constants
appearing in the original interaction Lagrangian. Writing out explicitly only the scalar,
pseudoscalar and vector terms, which are relevant
for our present discussion, we have
\begin{equation}
{\cal L}={\overline \psi}\left(i \fslash{\partial}-m\right) \psi + G_{\pi}
\left( \left(\overline{\psi}\psi \right)^2
	- \left(\overline{\psi}(\gamma_5\bold{\tau})\psi\right)^2 \right) 
- G_{\omega} \left( \overline{\psi} \gamma^{\mu} \psi \right)^2  + \ldots
\label{lag}
\ee
where $m$ is the current quark mass.
In the nuclear medium characterized by the density $\rho$, the quark bilinears 
$\overline{\psi}\psi$ and $\overline{\psi}\gamma^{\mu}\psi$ have expectation
values which we will separate as usual according to
$\overline{\psi}\psi=\langle \rho|\overline{\psi}\psi|\rho \rangle + (:\overline{\psi}\psi:)$
and $\overline{\psi}\gamma^{\mu}\psi=\langle \rho|\overline{\psi}\gamma^{\mu}\psi|\rho \rangle 
+ (:\overline{\psi}\gamma^{\mu}\psi:)$. The Lagrangian can then be expressed as
\be
{\cal L}={\overline \psi}\left(i \fslash{\partial}- M - 2 G_{\omega} \gamma^{\mu} \omega_{\mu} 
\right) \psi - \frac{(M-m)^2}{4 G_{\pi}} + G_{\omega} \omega_{\mu}\omega^{\mu} + {\cal L}_I,
\label{lag1}
\ee
where we defined $M=m-2G_{\pi} \langle \rho|\overline{\psi}\psi|\rho \rangle$ and
$\omega^{\mu}= \langle \rho|\overline{\psi}\gamma^{\mu}\psi|\rho \rangle$, and ${\cal L}_I$
is the normal ordered interaction Lagrangian\footnote{Our conventions are as follows: 
In the effective potential, $M$ is considered as 
a variable, and its 'physical' value is denoted as $M^*$ (solution of the gap equation in the medium)
or $M_0$ (solution of the gap equation in the vacuum). The quantity
$\langle \rho|\overline{\psi}\psi|\rho \rangle$ is considered as a function of $M$,
and its value at density $\rho$ and $M=M^*$ is denoted as 
$\langle \overline{\psi}\psi \rangle^*$, while its value for zero density and $M=M_0$
is the physical vacuum value $\langle \overline{\psi}\psi \rangle _0$. 
Concerning the diquark and nucleon masses $M_D \equiv M_D(M)$ and $M_N \equiv M_N(M)$, 
their values at $M=M^* 
\,\,(M=M_0)$ are denoted as $M_D^*$ and
$M_N^*$ ($M_{D0}$ and $M_{N0}$). Similar notation is used later for the energy density,
${\cal E}$, the nucleon energy, $\epsilon(k)$, the forward scattering amplitude, 
$f({\bold k}',{\bold k})$, the $\pi N$ sigma term, $\Sigma_{\pi N}$, etc.
(E.g; ${\cal E}^*={\cal E}(M=M^*)$, etc.) Further, all four dimensional momentum integrals, like 
that in Eq.(\ref{bubbs}), stand
symbolically for their regularized expressions in one of the schemes considered in this paper
(Euclidean cut-off, 3-momentum cut-off, or proper time cut-off scheme). We also note that
${\cal L}_I$ in (\ref{lag1}) also contains the counter terms linear in the normal ordered
products.}.

We briefly recapitulate the procedure to 
construct the nucleon as a quark-diquark state
for zero baryon density. Using a further Fierz transformation, one can decompose the
interaction Lagrangian in Eq.(\ref{lag1}) into a sum of qq channel interaction terms
\cite{FAD1}. For
our purposes we need the term which describes the qq interaction in the scalar diquark
($J^{\pi}=0^+, T=0$, color ${\overline 3}$) channel:
\be
{\cal L}_{I,s} = G_s \left(\overline{\psi}\left(\gamma_5 C\right)\tau_2
\beta^A \overline{\psi}^T\right) \left(\psi^T\left(C^{-1}\gamma_5\right)
\tau_2 \beta^A \psi\right),
\label{lags}
\ee
where $\beta^A=\sqrt{3/2}\,\, \lambda^A \,\,(A=2,5,7)$ are the color 
${\overline 3}$ matrices, and $C=i\gamma_2 \gamma_0$. The coupling
constant $G_{s}$ is a function of the coupling constants appearing in
the original interaction Lagrangian. The reduced t-matrix in the scalar
diquark channel is then obtained from the Bethe-Salpeter (BS) equation as
\be
\tau_{s}(q)={4iG_s \over 1 + 2 G_s \Pi_s(q)}
\label{taus} 
\ee
with the scalar qq bubble graph
\be
\Pi_s(q)=6 i \kint{k} tr_D \biggl[
	\gamma_5 S(k) \gamma_5  S\left(-(q-k)\right)\biggr].
\label{bubbs}
\ee
Here 
\be
S(q) = {1 \over \fslash{q} - M +i\epsilon}
\label{con}
\ee
is the Feynman propagator for the constituent quark, and $tr_D$ denotes the trace
over the Dirac indices. If one restricts the interacting qq channels to the
scalar one,
the relativistic Faddeev equation \cite{FAD1} can be reduced to
an effective BS equation for a composite scalar diquark and a quark
interacting
via quark exchange \cite{CAH}. This has been solved numerically by using the
Euclidean sharp cut-off scheme. For finite density, however, the gap equation (to be given
below) depends
on the nucleon mass, and in order to find its solution one has to 
solve the Faddeev equation many times. In this
paper, as a first step, we will restrict ourselves to the static approximation 
\cite{STAT1,STAT2} to the Faddeev equation, where 
the Feynman propagator in the quark exchange kernel
is simply replaced by $-1/M$. Then the Faddeev
equation reduces to a geometric series of quark-diquark bubble graphs
($\Pi_N(p)$), and the solution for the t-matrix in the color singlet
channel is
\be
T(p)=\frac{3}{M} \frac{1}{1+\frac{3}{M} \Pi_N(p)}
\label{tn}
\ee
with
\be
\Pi_N(p)=-\kint{k} S(k) \tau_s(p-k).
\label{bubbn}
\ee
The nucleon mass $M_N$ is then obtained as a solution of the equation \\ 
${\displaystyle 1+\frac{3}{M} \Pi_N(\fslash{p}=M_N)=0}$. 

\subsection{A hybrid model for the nuclear matter ground state}
Using hadronization and path integral techniques, any quark Lagrangian of the NJL type
can be expressed in terms of physical hadron degrees of freedom (nucleons
and mesons), see Ref.\cite{REIN,FAD2}. The result is a complicated 
non-local Lagrangian
of the linear $\sigma$ model type, which also contains interaction
terms of higher order than in the standard $\sigma$ model. Since in the
course of the derivation one has to integrate over the quark fields as well
as the auxiliary diquark and color octet baryon fields, ``trace log terms'' also appear
in the Lagrangian and these should be taken into account in the
computation of the effective potential. The resulting $\sigma$ model type
Lagrangian can then, in principle, be applied to nuclear matter by introducing
a chemical potential for the nucleons and using some standard approximation
scheme like the mean field approximation. 

In this paper, however, we will not fully exploit this ambitious method, but we will
resort to the most simple, nontrivial approximation. First, in the ``trace
log terms'' we take into account only the quark loop term, that is we
neglect the diquark loop and baryon octet loop terms which can be thought
of as ring type quark-quark and quark-diquark correlations in the Dirac sea of quarks.
Second, the resulting effective $\sigma$ model-type Lagrangian is treated on the mean field
level. In this way one obtains the following expression for the effective
potential (energy density)\footnote{In this paper we will frequently refer to the
function ${\cal E}$ as the ``effective potential'', even though we use the 
density as a variable rather
than the chemical potential.}:
\be
{\cal E}={\cal E}_V -G_{\omega} \omega_{\mu}\omega^{\mu} + \gamma_N \int 
\frac{d^3 k}{(2 \pi)^3} n(k) \epsilon(k),
\label{en}
\ee
where the contribution due to the quark loop is
\be
{\cal E}_V=i\,\gamma_q \kint{k} \, {\rm ln}\, \, \frac{k^2-M^2+i\epsilon}{k^2-M_0^2+i\epsilon}
+\frac{(M-m)^2}{4 G_{\pi}} -\frac{(M_0-m)^2}{4 G_{\pi}}
\label{env}       
\ee
with $\gamma_q=12$.
In Eq.(\ref{en}), $n(k)$ is the Fermi distribution function for the nucleons,
$\gamma_N=4$, and $\epsilon(k)$ is the energy of the nucleon (pole of the
quark-diquark t-matrix) moving in the scalar and vector fields which couple
to the quarks.  

This simple model can also be motivated by defining a ``hybrid approximation'' for
the nuclear matter ground state as follows. The nuclear matter expectation 
value of any local quark operator, ${\cal O}$, consists of its expectation
value in the ``valence quark vacuum'', $|0\rangle = |\rho=0>$, 
and an average over the nucleon Fermi sea consisting of correlated valence ($v$) quarks:
\be
\langle \rho|{\cal O}|\rho \rangle = \langle 0|{\cal O}|0 \rangle +    
\gamma_N \int \frac{d^3 k}{(2 \pi)^3} n(k) \langle{\cal O}\rangle_v({\bold k}),
\label{hyb}
\ee
where 
\begin{equation}
\langle{\cal O}\rangle_v({\bold k})\equiv
\int d^3 r\left[\langle N,{\bold k}|{\cal O}({\bold r})|N,{\bold k}\rangle - 
\langle 0|{\cal O}({\bold r})|0 \rangle\right]
\end{equation}
with 
$|N,{\bold k}\rangle$ denoting the correlated valence quark state (quark-diquark state in
our case) of the nucleon. (We divided throughout by an
overall volume factor $V$ in Eq.(\ref{hyb}).)
Taking ${\cal O}={\cal H}$, where ${\cal H}$ is the quark Hamiltonian density corresponding
to the Lagrangian density (\ref{lag1}), we have $\langle{\cal H}\rangle_v({\bold k})
=\epsilon(k)$, and $\langle 0|{\cal H}|0 \rangle$ in the mean field approximation 
is given
by ${\cal E}_V-G_{\omega}\omega_{\mu}\omega^{\mu}$ of Eq.(\ref{en}).
In this way one arrives at Eq.(\ref{en}). We will see
later that, for ${\cal O}={\overline \psi}\psi$ or ${\cal H}$, Eq.(\ref{hyb}) is consistent
with the well known low density expansions of the quark condensate and the energy density.

\subsubsection{Mean vector field}
Let us first discuss the form of $\epsilon({\bold k})$ and determine the mean vector field:
As is clear from Eq.(\ref{lag1}), the constituent quark propagator in the
scalar and vector fields is obtained by replacing $k^{\mu} \rightarrow k_Q^{\mu} 
\equiv k^{\mu}-2 G_{\omega} \omega^{\mu}$ in the propagator (\ref{con}). 
Therefore, the qq bubble graph in the mean field approximation is obtained from (\ref{bubbs})
by replacing $S(k)\rightarrow S(k_Q)$ and $S(-(q-k))\rightarrow S(-(q-k)_Q)=S(k_Q-q_D)$,
where $q_D^{\mu}\equiv q^{\mu}-4G_{\omega}\omega^{\mu}$. Thus the bubble graph
and the reduced t-matrix in the mean fields are obtained by replacing $q^{\mu}\rightarrow q_D^{\mu}$
in (\ref{bubbs}) and (\ref{taus}), respectively. Consequently, the quark-diquark bubble graph
in the mean fields is obtained from (\ref{bubbn}) by replacing $S(k)\rightarrow S(k_Q)$
and $\tau(p-k)\rightarrow \tau((p-k)_D)=\tau(p_N-k_D)$, where 
\begin{equation}
p_N^{\mu} \equiv p^{\mu}-6G_{\omega}\omega^{\mu}. 
\end{equation}
As a result, the quark-diquark bubble graph and the quark-diquark
t-matrix in the mean fields is obtained by replacing $p^{\mu}\rightarrow p_N^{\mu}$ 
in (\ref{bubbn})
and (\ref{tn}), respectively. The spectrum is then obtained by noting that if the equation
$1+\frac{3}{M}\Pi_N(p)=0$ is satisfied at $\fslash{p}=M_N$, the solution of 
$1+\frac{3}{M}\Pi_N(p_N)=0$ is $\fslash{p_N}=M_N$, and solving for $p^0=\epsilon(p)$ 
this gives the positive energy pole as
\be
\epsilon(p)=6 G_{\omega} \omega^0 + 
\sqrt{{\bold p}_N^2 + M_N^2} \equiv 6 G_{\omega} \omega^0 + 
E_N(p_N).
\label{spec}
\ee
Using this in Eq.(\ref{en}) we obtain the physical vector field ($\omega^{*\mu}$) from 
the condition 
$\left(\partial {\cal E}/\partial \omega^{\mu}\right)_{\omega=\omega^*}=0$ as 
\begin{equation}
\omega^{*\mu}=3 j_B^{\mu}, 
\end{equation}
where
the baryon current is given by 
\begin{equation}
j_B^0=\rho,\,\,\,\,\,\,\,\,\,\, 
{\bold j}_B=\gamma_N \int \frac{d^3 k}{(2 \pi)^3} n(k) \frac{{\bold k}_N}{E_N(k_N)}.
\end{equation}
Using this to eliminate
the vector field, we obtain for the energy density
\be
{\cal E}={\cal E}_V + {\cal E}_{\omega} + {\cal E}_F,
\label{en1}
\ee
where the vacuum part has been given in Eq.(\ref{env}), and 
\begin{eqnarray}
{\cal E}_{\omega} &=& 9 G_{\omega} \left(\rho^2+{\bold j}_B^2\right), \\
\label{enw}
{\cal E}_F &=& \gamma_N \int \frac{d^3 k}{(2 \pi)^3} n(k) E_N(k_N).
\label{enf}
\end{eqnarray}
Unless stated explicitly, we will consider nuclear matter at rest, where ${\bold \omega}^*={\bold j}_B=0$, 
${\bold k}_N={\bold k}$, and $n(k)$ becomes the usual step function $\Theta(p_F-k)$.

\subsubsection{Mean scalar field}
The mean scalar field in the medium (or the constituent quark mass) is determined by the condition 
\be
\partial {\cal E}/\partial M = \frac{M-m}{2G_{\pi}} +  
\langle \rho|{\overline \psi}\psi|\rho\rangle =0,
\label{dcond}
\ee
where the second equality is the gap equation which determines the physical value
of $M$ ($M=M^*$). 
The quark condensate in (\ref{dcond}) is given by (see Eq.(\ref{hyb}))
\be
\langle \rho|{\overline \psi}\psi|\rho\rangle = \langle 0|{\overline \psi}\psi|0\rangle
+ \gamma_N \int \frac{d^3 k}{(2 \pi)^3} n(k) \langle {\overline \psi}\psi\rangle_v(k)
\label{hybc}
\ee
with
\begin{eqnarray}
\langle 0|{\overline \psi}\psi|0\rangle &=& -2i \gamma_q M \kint{k} \frac{1}{k^2-M^2+i\epsilon}  
\label{conv} \\
\langle {\overline \psi}\psi\rangle_v(k) &=& \frac{\partial 
\epsilon(k)}{\partial M}
=\frac{M_N}{E_N(k)}\frac{\partial M_N}{\partial M} \label{cond}.
\end{eqnarray}
We finally note the following relation, which follows from (\ref{dcond}):
\begin{eqnarray}
2 G_{\pi} \frac{{\rm d} M^*}{{\rm d} m} &=& \frac{2G_{\pi}}{1+2G_{\pi} 
\left(\frac{\partial}{\partial M}
\langle 0|{\overline \psi}\psi|0\rangle\right)_{M=M^*}} \label{rel1} \\ 
&=& \left(\frac{\partial^2 {\cal E}}{\partial M^2}\right)^{-1}_{M=M^*} 
\equiv \frac{g^2}{M_{\sigma}^{*2}}. 
\label{rel2}
\end{eqnarray}
Here $g$ is the meson-quark coupling constant at zero density and zero meson momentum\footnote
{The meson-quark coupling constant $g$ is introduced such that, in terms of the 
normalized meson fields 
$\sigma=-\frac{2G_{\pi}}{g} {\overline \psi}\psi$ and ${\bold \pi}=-\frac{2G_{\pi}}{g} 
{\overline \psi} i \gamma_5 {\bold \tau}\psi$, the Yukawa coupling terms in the semi-bosonized 
Lagrangian read $-{\overline \psi}
g\left(\sigma + i{\bold \pi}\cdot {\bold \tau} \gamma_5\right)\psi$. In terms of the pionic bubble
graph in the vacuum $(\Pi_{\pi V}(k^2))$ it is given by $g^2=-1/\frac{\partial \Pi_{\pi V}}
{\partial k^2}$ at $k^2=0$ and $M=M_0$. We also note that for $m=0$ we have the relations $M_0 g = f_{\pi}$
and $M_{\sigma 0}=2M_0$ from the gap equation in the vacuum and the $q {\overline q}$ bubble graph
for pion decay.}, and Eq.(\ref{rel2}) defines  
the effective $\sigma$ mass in the medium, $M_{\sigma}^*\equiv M_{\sigma}(M=M^*)$, 
in terms of the curvature of the effective
potential. 
The zero density analogs to (\ref{rel1}), (\ref{rel2}) are obtained by replacing 
$M^*\rightarrow M_0$, ${\cal E}\rightarrow {\cal E}_V$, $M_{\sigma}^2\rightarrow
M_{\sigma 0}^2$. 

\subsection{Expansions in powers of the density}
In this subsection we derive various model independent expansions in powers of the density,
as well as some relations which are peculiar to the hybrid model introduced in the
previous subsection. These discussions will help to
understand the physics behind the hybrid model expressions (\ref{en}) or (\ref{hyb}), in
particular the role of the vacuum terms. They 
will also be useful when we discuss the numerical results in sect.3, in particular the
mechanism for saturation. 

The validity of the Feynman-Hellman theorem
\be
\langle {\overline \psi}\psi\rangle^* - \langle {\overline \psi}\psi\rangle_{0}= 
\frac{d\,{\cal E}^*}{d\,m}.
\label{fh}
\ee
in the present hybrid model is easily seen as follows.
Relation (\ref{dcond}) at the ``physical value'', $M=M^*$, reads 
$M^*=m-2G_{\pi}\langle {\overline \psi}\psi
\rangle^*$ (see the first footnote in this section for our notation). 
From this we can express the
difference between the condensate in the medium and in the vacuum as $\langle {\overline \psi}\psi
\rangle^*-\langle {\overline \psi}\psi\rangle_{0}=-(M^*-M_0)/2 G_{\pi}$. On the other hand, 
since the
energy density is stationary at $M=M^*$ we also have the relation $d\,{\cal E}^*/d\,m = 
\partial {\cal E}^*/\partial m = -(M^*-M_0)/2 G_{\pi}$, where in the last equality we 
used the form (\ref{en}). This leads to the Feynman-Hellman theorem (\ref{fh}).

The expansion of the energy density ${\cal E}^*={\cal E}(M=M^*)$ around some fixed value
of the density can be derived in a general
(model independent) way by applying Landau's expression for $\delta {\cal E}^*$ to the
case of spin-isospin 
independent variations of the nucleon distribution functions, $\delta n_k$, caused by
a deviation of the Fermi momentum from some value $p_{F0}$ (density $\rho_0$):
\be
\delta {\cal E}^*=\gamma_N \int \frac{d^3 k}{(2\pi)^3} \epsilon^*(k;\rho_0) \delta n_k 
+ \frac{1}{2}\gamma_N \int \frac{d^3 k}{(2\pi)^3} \gamma_N \int \frac{d^3 k'}{(2\pi)^3}
f^*({\bold k}',{\bold k};\rho_0) \delta n_{k'} \delta n_k +\dots
\label{landau}
\ee
The Landau-Migdal interaction refers to the case $|{\bold k}'|=
|{\bold k}|=p_{F}$ and is denoted as $f^*({\rm cos} \,\Theta)$, where $\Theta$ is the
angle between ${\bold k}'$ and ${\bold k}$. We insert
\be
\delta n_k = \sum_{n=1}^{\infty} \frac{(p_F-p_{F0})^n}{n!} (-1)^{n-1} 
\frac{\partial^{n-1}}{\partial k^{n-1}} \delta(p_F-k)
\label{var}
\ee
into Eq.(\ref{landau}). First we derive the low density expansion and set $p_{F0}=0$. 
Note that each phase space factor
in (\ref{landau}) contains a factor $k^2$, so that the first term of (\ref{landau}) 
involves $p_F^3$ and the second term $p_F^6$. The first term is simply the
expansion of the energy density of a free Fermi gas (${\cal E}_0$) in powers of 
$p_F$. Performing partial integrations for the second term we obtain the low density
expansion
\be
{\cal E}^*=\rho \left(M_{N0} + \frac{3}{10} \frac{p_F^2}{M_{N0}}\right) 
+ \frac{\rho^2}{2} f_{0,L=0} 
+ {\cal O}(p_F^7),
\label{lowd}
\ee 
where the Landau-Migdal parameter $f^*_{L=0}$ is defined as $f^*_{L=0}=\frac{1}{2}
\int_{-1}^{1} {\rm d\,cos} \Theta \\
f^*({\rm cos} \Theta)$, and $f_{0, L=0}$ is the zero density value 
\footnote{The relation to the spin averaged s-wave scattering length in the vacuum ($a_{L=0}$) is 
$f_{0,L=0}=\frac{4 \pi}{M_{N0}} a_{L=0}$. In order to obtain the low density
expansion, we assumed the Landau-Migdal interaction to be finite. The contribution of the 
pionic Fock
term is proportional to the average over the Landau angle of 
${\bold q}^2/({\bold q}^2+M_{\pi0}^2)^2$
with ${\bold q}^2=2p_F^2(1-{\rm cos}\Theta)$. If one sets $M_{\pi 0}=0$ from the outset,
this goes as $1/p_F^2$, and its contribution to the energy density and the quark
condensate would be $\propto \rho^{4/3}$ instead of $\rho^2$ \cite{QCD}. However, in the real
world the low density expansion always implies $p_F<<M_{\pi 0}$. We also note that the origin
of the $p_F^7$ term in (\ref{lowd}) is, besides the relativistic correction to the free Fermi
gas energy density, the three-body term which is the next term in the
expansion (\ref{landau}). The three-body reducible piece of this term behaves as $1/p_F^2$
as $p_F\rightarrow 0$, and therefore gives rise to a contribution $\propto a_{L=0}^2 \,\,p_F^7$
\cite{FW}.}.        

The corresponding low density expansion of the quark condensate is then obtained from
the Feynman-Hellman theorem (\ref{fh}) as
\be
\langle {\overline \psi}\psi \rangle^* - \langle {\overline \psi}\psi \rangle_0
= \rho\,\frac{{\rm d}M_{N0}}{{\rm d}m}\left(1-\frac{3}{10} \frac{p_F^2}{M_{N0}^2}\right)
+\frac{\rho^2}{2}\frac{{\rm d}f_{0,L=0}}{{\rm d}m} + {\cal O}(p_F^7).
\label{lowc1}
\ee
We introduce the following definitions of the one and two body $\pi N$ ``$\sigma$-terms'', 
which we
give here for the case of finite density for later use\footnote{Our notation in 
(\ref{sigma12}) indicates that the quantity $\Sigma_{\pi N}^{'*}$ is the
derivative of the momentum dependent one-body $\sigma$-term at finite density
($\Sigma_{\pi N}^*(k) = m\,\frac{{\rm d}\epsilon^*(k)}{{\rm d}m}$) at $k=p_F$.
We remark that, to our knowledge, the $\pi NN$ sigma term of Eq. (\ref{sigma12}) is
introduced here for the first time. It is the expectation value of $\overline{\psi} \psi$
between a correlated two-nucleon state, and it would be interesting to investigate 
whether there is a relation to observables of pion-deuteron scattering.}:
\be
\Sigma_{\pi N}^* \equiv m\,\frac{{\rm d}\epsilon^*_F}{{\rm d}m};
\,\,\,\,\,\,\,\,\,\,\Sigma_{\pi N}^{'*} \equiv m\,\frac{{\rm d}v_F^*}{{\rm d}m};   
\,\,\,\,\,\,\,\,\,\,
\Sigma_{\pi NN}^* \equiv m\,\frac{{\rm d}f_{L=0}^*}{{\rm d}m} \label{sigma12},
\ee
where $\epsilon_F^*$ is the Fermi energy and 
$v_F^*=\left({\rm d}\epsilon^*(k)/{\rm d}k\right)_{k=p_F}$ is the 
Fermi velocity. For zero density, the first relation in (\ref{sigma12}) 
reduces to the usual definition of 
the one body $\pi N$ $\sigma$-term ($\Sigma_{\pi N,0}= 
m\frac{{\rm d}M_{N0}}{{\rm d}m}$).
The Gell-Mann-Oakes-Renner (GOR) relation in the NJL model reads \cite{LIGHT}
$m\,\langle {\overline \psi}\psi \rangle_0 = -(F_{\pi}M_{\pi 0})^2\,C$ with
$C=\left(1-\frac{m}{M_0}\right)$, where we define the pion decay constant, $F_{\pi}$, 
and pion mass 
in the vacuum, $M_{\pi 0}$, at zero momentum instead of at the pion pole.  
Then Eq.(\ref{lowc1}) can be written in the following form:
\be
\frac{\langle {\overline \psi}\psi \rangle^*}{\langle {\overline \psi}\psi \rangle_0}
= 1 - \rho\,\frac{\Sigma_{\pi N,0}}{(F_{\pi}M_{\pi 0})^2 C}
\left(1-\frac{3}{10} \frac{p_F^2}{M_{N0}^2}\right) - \frac{\rho^2}{2}
\frac{\Sigma_{\pi NN,0}}{(F_{\pi}M_{\pi 0})^2 C} + {\cal O}(p_F^7).
\label{lowc2}
\ee    

All relations given above are model independent. It is a simple task to check their
validity in our hybrid model, and to give the corresponding low density expansions
of the constituent quark mass and the effective nucleon mass:   
Starting from the expression (\ref{en1}) for the energy density and using the fact that
the vacuum part ${\cal E}_V$ is stationary at $M=M_0$, we obtain the expansion
\be
{\cal E}(M)-{\cal E}_0
= \frac{(M-M_0)^2}{4 G_{\pi}\frac{\partial M_0}{\partial m}} + \rho \, (M-M_0) 
\frac{\partial M_{N0}}{\partial M_0} + 9 G_{\omega} \rho^2 + \dots,
\label{expe1}
\ee 
where we used (\ref{rel1}), (\ref{rel2}) in the vacuum.  
Minimizing (\ref{expe1}) with respect to $M$ we obtain $M^*-M_0=-2 G_{\pi}\rho
\frac{{\rm d}M_{N0}}{{\rm d}m}+\dots$, which is the lowest order term in (\ref{lowc1}), and inserting
this into the above expression gives
\begin{eqnarray}
{\cal E}^*-{\cal E}_0 &=& -\frac{1}{2} \rho^2 \left(\frac{\partial M_N}{\partial M}\right)^2_{M=M_0}
\,\left(\frac{\partial^2 {\cal E}_V}{\partial M^2}\right)^{-1}_{M=M_0} + 
9 G_{\omega} \rho^2 + \dots
\label{expe2} \\
&\equiv& \frac{1}{2} \rho^2 \,\left(-\frac{g_{\sigma 0}^2}{M_{\sigma 0}^2}
+\frac{g_{\omega}^2}{M_{\omega}^2}\right) + \dots \label{expe3}
\end{eqnarray}
which is Eq.(\ref{lowd}) with the scattering length given by $\sigma$ and $\omega$
exchange between the nucleons. The factor $\frac{1}{2}$ arises here because the vacuum 
part (first term in (\ref{expe1})) cancels half of the attractive Fermi part (second term in (\ref{expe1})). 
The $\sigma$ 
meson mass has already been defined in terms of the curvature of the effective
potential in (\ref{rel2}), and $g_{\sigma 0}$ is the vacuum (zero density) value of the 
$\sigma NN$ coupling
constant at zero momentum, $g_{\sigma}^*$, which is defined by 
\be
\left(\frac{\partial M_{N}}{\partial M}\right)_{M=M^*} = \frac{g^*_{\sigma}}{g}.
\label{coupls}
\ee
(The $\sigma qq$ coupling constant at zero density, $g$, has been introduced in sect. 2.1.2.)   
For the $\omega$ meson exchange part in (\ref{expe3}), we identified
\be
G_{\omega}=\frac{1}{18} \frac{g_{\omega}^2}{M_{\omega}^2}.    
\label{couplw}
\ee

Since we have already confirmed the validity of the Feynman-Hellman theorem ({\ref{fh})
in our hybrid model, the relations (\ref{lowc1}) and (\ref{lowc2}) are also valid\footnote{It 
is, of course, also possible to derive (\ref{lowc1}) directly from the
hybrid model relation (\ref{hybc}). For example, the term $\propto \rho$ is easily
obtained by expanding (\ref{hybc}) as follows:
\begin{eqnarray}
\langle {\overline \psi}\psi\rangle^* = \langle {\overline \psi}\psi\rangle_{0} + (M^*-M_0)
\left(\frac{\partial \langle 0|{\overline \psi}\psi|0\rangle}{\partial M}\right)_{M=M_0} + 
\rho \left(\frac{\partial M_{N}}{\partial M}\right)_{M=M_0} + \dots
\nonumber
\end{eqnarray}
By using $M^*-M_0=-2G_{\pi}(\langle {\overline \psi}\psi\rangle^* - 
\langle {\overline \psi}\psi\rangle_0)$ and Eq.(\ref{rel1}) we easily obtain the
linear term in (\ref{lowc1}).}.
The corresponding expansions of the constituent quark mass and the effective nucleon mass in
our model are then easily obtained as
\begin{eqnarray}
\frac{M^*}{M_0}
&=& 1 - \rho\,\frac{\Sigma_{\pi N,0}}{(F_{\pi}M_{\pi 0})^2}
\left(1-\frac{3}{10}\frac{ p_F^2}{M_{N0}^2}\right) - \frac{\rho^2}{2}
\frac{\Sigma_{\pi NN,0}}{(F_{\pi}M_{\pi0})^2}  \nonumber \\
&+& {\cal O}(p_F^7) \label{lowmq} \\
\frac{M_N^*}{M_{N0}}
&=& 1 - \rho\,A_0\,\frac{\Sigma_{\pi N,0}}{(F_{\pi}M_{\pi0})^2}
\left(1-\frac{3}{10}\frac{ p_F^2}{M_{N0}^2}\right) - \frac{\rho^2}{2} \nonumber \\
&\times& \left(A_0\,\frac{\Sigma_{\pi NN,0}}{(F_{\pi}M_{\pi0})^2}
-B_0\,\left(\frac{\Sigma_{\pi N,0}}{(F_{\pi}M_{\pi0})^2}\right)^2 \right) 
+ {\cal O}(p_F^7),
\label{lowmn}
\end{eqnarray}
where in (\ref{lowmn}) we introduced the zero density values of the following quantities:
\be
A^*=\left(\frac{\partial M_N}{\partial M}\right)_{M=M^*}\frac{M_0}{M_{N0}},
\,\,\,\,\,\,\,\,\,\,\,\,\,\,\,\,\,\,\,\,
B^*=\left(\frac{\partial^2 M_N}{\partial M^2}\right)_{M=M^*}\frac{M_0^2}{M_{N0}}.
\label{defab}
\ee
The ``naive'' ($M_N=3M$) values are $A_0=A^*=1$ and $B_0=B^*=0$, and for this case the term
proportional to $\rho$ in (\ref{lowmn}) agrees with the result given by Eq.(5.16) of 
Ref.\cite{QCD1}.

In a similar way one can expand
the energy density, quark condensate, etc, around any density ($\rho_0>0$). The
difference is that such an expansion in powers of $(\rho-\rho_0)$ is analytic, while
the low density expansion discussed above is analytic in the Fermi momentum rather
than the density. Therefore, in order the get the expansion of ${\cal E}^*$ up to second order
in $(\rho-\rho_0)$, it is sufficient to consider the $n=1$ and $n=2$ terms in
(\ref{var}) to obtain the well know expression
\be
{\cal E}^*(\rho)-{\cal E}^*(\rho_0) = (\rho-\rho_0) \epsilon^*_F(\rho_0)  + \frac{1}{2}
(\rho-\rho_0)^2 \left(\frac{\pi^2}{2 p_{F0}^2} v^*_F(\rho_0) + f^*_{L=0}(\rho_0) \right)
+{\cal O}\left((\rho-\rho_0)^3\right).
\label{exen}
\ee
Using the Feynman-Hellman theorem (\ref{fh}), this can be converted into the
corresponding expansions of the quark condensate, and in our hybrid model this in turn
gives the expansions of the constituent quark mass and
the nucleon mass around any density $\rho_0$: 
\begin{eqnarray} 
& & \frac{\langle {\overline \psi}\psi\rangle^*(\rho) - \langle {\overline \psi}\psi\rangle^*(\rho_0)}
{\langle {\overline \psi}\psi\rangle_{0}} = 
\frac{-1}{(F_{\pi}M_{\pi0})^2 C}\left\{(\rho-\rho_0)\Sigma_{\pi N}^*(\rho_0)
+ \frac{1}{2} (\rho-\rho_0)^2 \right. \nonumber \\
&\times& \left. \left(\frac{\pi^2}{2p_{F0}^2} 
\Sigma_{\pi N}^{'*}(\rho_0) + \Sigma_{\pi NN}^*(\rho_0)\right)\right\}
\label{exdco} \\
& & \frac{M^*(\rho) - M^*(\rho_0)}{M_0} = 
\frac{-1}{(F_{\pi}M_{\pi0})^2 }\left\{(\rho-\rho_0)\Sigma_{\pi N}^*(\rho_0)
+ \frac{1}{2} (\rho-\rho_0)^2 \right. \nonumber \\
&\times& \left. \left(\frac{\pi^2}{2p_{F0}^2} 
\Sigma_{\pi N}^{'*}(\rho_0) + \Sigma_{\pi NN}^*(\rho_0)\right)\right\} 
\label{exdmq} \\
& & \frac{M^*_N(\rho) - M^*_N(\rho_0)}{M_{N0}} = 
\frac{-1}{(F_{\pi}M_{\pi0})^2 }\left\{(\rho-\rho_0) A^*(\rho_0)\Sigma_{\pi N}^*(\rho_0)
+ \frac{1}{2} (\rho-\rho_0)^2 \right. \nonumber \\
&\times& \left. \left[A^*(\rho_0)\left(\frac{\pi^2}{2p_{F0}^2} 
\Sigma_{\pi N}^{'*}(\rho_0) + \Sigma_{\pi NN}^*(\rho_0)\right) 
-  B^*(\rho_0)\left(\frac{\Sigma_{\pi N}^*(\rho_0)}
{(F_{\pi} M_{\pi0})} \right)^2\right]\right\}. \nonumber \\
\label{exdmn}
\end{eqnarray}
We finally note for later use that in our model the one-body $\sigma$-term at finite density can be
expressed as (see eqs.(\ref{spec}), (\ref{rel1}), (\ref{rel2}) and (\ref{coupls})):
\be
\Sigma^*_{\pi N}=m \frac{{\rm d}\epsilon^*_F}{{\rm d}m}=
m \frac{M_N^*}{E_F^*} \left(\frac{\partial M_N}{\partial M}\right)_{M=M^*} 
\frac{{\rm d} M^*}{{\rm d} m} = 
\frac{m}{2 G_{\pi}} \frac{M_N^*}{E_F^*}\frac{g g^*_{\sigma}}{M^{*2}_{\sigma}},
\label{onesig}
\ee
and therefore the derivative of the constituent quark mass w.r.t. the density is determined
by the $\sigma$ meson exchange interaction between a quark and a nucleon, while  
the derivative of the effective nucleon mass is determined by the nucleon-nucleon interaction
due to $\sigma$ meson exchange. (Note the additional factor $A^*\propto g^*_{\sigma}/g$ in (\ref{exdmn}).)
This point will be discussed further in the next subsection.


\subsection{The Landau-Migdal interaction and a mechanism for saturation
in chiral Hartree-type theories}
In this section we derive the form of the effective two-nucleon interaction (forward scattering
amplitude) from the definition
(\ref{landau}), and discuss some important features which arise from the quark substructure
of the nucleon. For this purpose, we follow closely
the method explained in Ref.\cite{BLA}. Since the energy density is stationary at
$X=X^*$, where $X$ denotes both $M$ and $\omega^{\mu}$, the quasiparticle energy 
is given by\footnote{Here the label $i$ denotes a quasiparticle state. We use the symbol 
$\delta$ for
total variations, and $\partial$ for partial variations.}
\be
\epsilon^*_i=\frac{{\delta}{\cal E}^*}{{\delta}n_i} = \frac{\partial {\cal E}^*}{\partial n_i},
\label{qen}
\ee 
and the effective two-nucleon interaction by 
\be
f^*_{ij}=\frac{{\delta}^2{\cal E}^*}{{\delta}n_j{\delta}n_i} = 
\frac{{\delta}\epsilon^*_i}{{\delta}n_j} =
\frac{\partial \epsilon^*_i}{\partial n_j} + \left(\frac{\partial \epsilon_i}
{\partial M}\right)_{X=X^*} \frac{\partial M^*}{\partial n_j} + 
\left(\frac{\partial \epsilon_i}{\partial \omega_{\mu}}\right)_{X=X^*} \frac{\partial {\omega}^{*\mu}}
{\partial n_j}.
\label{lm}
\ee
Here we use the relations
\begin{eqnarray}
\frac{{\delta}}{{\delta}n_j} \left(\frac{\partial {\cal E}}{\partial X}\right)_{X=X^*} &=& 0 = 
\left(\frac{\partial^2 {\cal E}}{\partial n_j
\partial X}\right)_{X=X^*} + \left(\frac{\partial^2 {\cal E}}{\partial X \partial M}\right)_{X=X^*}
\frac{\partial M^*}{\partial n_j} \nonumber \\
&+& \left(\frac{\partial^2 {\cal E}}{\partial X \partial \omega_{\mu}}\right)_{X=X^*}
\frac{\partial \omega^{*\mu}}
{\partial n_j}.
\label{deriv}
\end{eqnarray}
In the limit of zero baryon current (nuclear matter at rest) 
we have \cite{BLA}
$\frac{\partial^2 {\cal E}}{\partial M \partial \omega_{\mu}}\rightarrow 0$, and we can 
use (\ref{deriv})
to eliminate $\frac{\partial M^*}{\partial n_j}$ and $\frac{\partial {\omega}^{*\mu}}
{\partial n_j}$ from Eq.(\ref{lm}). 
This gives the interaction in the form
\be
f^*_{ij}=\frac{\partial \epsilon^*_i}{\partial n_j} -  
\left[\left(\frac{\partial \epsilon_i}{\partial M}\right) 
\left(\frac{\partial^2 {\cal E}}{\partial M^2}\right)^{-1} \left(\frac{\partial \epsilon_j}
{\partial M}\right)
+ \left(\frac{\partial \epsilon_i}{\partial \omega_{\mu}}\right) 
\left(\frac{\partial^2 {\cal E}}{\partial \omega_{\mu} \partial \omega_{\nu}}\right)^{-1} 
\left(\frac{\partial \epsilon_j}{\partial \omega_{\nu}}\right)\right]_{X=X^*},
\label{lm1}
\ee
where the first term is zero in our hybrid model (see (\ref{spec}), and the other two terms 
have already the form of the direct 
$\sigma$ and $\omega$ exchange potentials with the meson masses given by the second 
derivatives of the energy density and the couplings to the nucleon by the derivatives
of the quasiparticle energy with respect to the mean fields (see eqs.(\ref{rel2}),
(\ref{coupls}) for the $\sigma$ meson). Carrying out the required derivatives by using (\ref{en})
and (\ref{spec}) and setting ${\bold j}_B=0$ at the end, we obtain
in our model for two nucleons with momenta ${\bold p}$, ${\bold p}'$:
\be
f^*({\bold p}',{\bold p}) = - \frac{M_N^{*2}}{E_N^*(p) E_N^*(p')} 
\frac{g_{\sigma}^{*2}}
{M_{\sigma V}^{*2} + \Pi_{\sigma D}^*} + \frac{g_{\omega }^2}{M_{\omega}^2} -
\frac{g_{\omega }^2}{M_{\omega}^2 + \frac{1}{2} g_{\omega }^2 \frac{\rho}{E_F^*}} 
\frac{{\bold p}'\cdot {\bold p}}{E_N^*(p') E_N^*(p)},
\label{landaum}
\ee
where $E_F=E_N(p_F)$.
The $\sigma$NN coupling constant is given by (\ref{coupls}), and the squared $\sigma$ mass 
has been split
into a vacuum part and a density part ($M_{\sigma}^2=M_{\sigma V}^2+\Pi_{\sigma D}$), 
which are given by
\begin{eqnarray}
M_{\sigma V}^2 &=& g^2 \left(\frac{1}{2 G_{\pi}} -2i\gamma_q \kint{k} \frac{k^2+M^2}{\left(k^2-M^2+i\epsilon\right)^2}
\right) \label{msigma} \\
\Pi_{\sigma D} &=& g_{\sigma }^2\, \gamma_N \int \frac{d^3 k}{\left(2\pi\right)^3} n(k) 
\frac{k^2}{E_N(k)^3}
+ g_{\sigma \sigma } \,\gamma_N \int \frac{d^3 k}{\left(2\pi\right)^3} n(k) \frac{M_N}{E_N(k)}.
\nonumber \\ 
\label{second}
\end{eqnarray}
The $\sigma \sigma$NN coupling constant in Eq.(\ref{second}) arises due to the 
curvature of the function $M_N(M)$ and is defined as
\be
\frac{g_{\sigma \sigma }}{g^2} \equiv \frac{\partial^2 M_N}{\partial M^2}.
\label{cont}
\ee
Concerning the $\omega$ meson exchange part in (\ref{landaum}), we used the definition
(\ref{couplw}). The last term in (\ref{landaum}) contributes to the Landau-Migdal
parameter $f^*_{L=1}$, and the first two terms give
\be
f^*_{L=0}=-\left(\frac{M_N^{*}}{E_F^{*}}\right)^{2} \,\frac{g_{\sigma }^{*2}}{M_{\sigma}^{*2}} + 
\frac{g_{\omega }^2}{M_{\omega}^2}.
\label{f0}
\ee 
This result, which we derived here for clarity directly from the definition of the
Landau-Migdal interaction, can of course also be obtained from the developments of the
previous subsection. From eqs. (\ref{exdmq}), (\ref{exdmn}) and (\ref{onesig}) we see that the derivative of the
effective quark and nucleon masses w.r.t. density can be expressed as
\begin{eqnarray}
\frac{\partial M^*}{\partial \rho} &=& \frac{M_N^*}{E_F^*}\left[\left(\frac{\partial M_N}{\partial M}\right)
\left(\frac{\partial ^2 {\cal E}}{\partial M^2}\right)^{-1}\right]_{M=M^*},
\label{dermqst} \\
\frac{\partial M_N^*}{\partial \rho} &=& \frac{M_N^*}{E_F^*}\left[\left(\frac{\partial M_N}{\partial M}\right)^2
\left(\frac{\partial ^2 {\cal E}}{\partial M^2}\right)^{-1}\right]_{M=M^*},
\label{dermst}
\end{eqnarray}
and therefore the derivative of the nucleon energy (\ref{spec}) at the Fermi surface is 
\be
\left(\frac{{\rm d} \epsilon^*(p)}{{\rm d} \rho}\right)_{p=p_F}=f^*_{L=0},
\ee
where $f^*_{L=0}$ is given by (\ref{f0}).

From the developments in this and the previous subsection we can make the following
important observation. If the function $M_N(M)$ for the
nucleon mass in the scalar field has an appreciable positive curvature 
$\partial^2 M_N / \partial M^2$
for large scalar potentials\footnote{Here and in the following, the ``scalar potential'' 
is defined as $M_0-M$.} 
(small $M$), (i.e., if $M_N(M)$ becomes ``flat'' for small $M$), 
there arise two 
important repulsive effects. First, the
$\sigma NN$ coupling constant (\ref{coupls}) is reduced for large scalar potentials 
(large densities). Second,
the sigma mass in the medium receives a repulsive contribution from the second term in
(\ref{second}), which can be expected to over-compensate the reduction of the first term at normal
densities, since the first term arises entirely from the Fermi motion and is of the order
$1/M_N^3$. Therefore, an appreciable
curvature of the function $M_N(M)$ could lead to a suppression of the attraction due to the
$\sigma$ meson exchange contribution to $f^*_{L=0}$, see (\ref{f0}). This could lead
to a scalar
potential for the nucleons which grows less rapidly with increasing density than the vector 
potential (which grows $\propto \rho$). If the r.h.s. of Eq.(\ref{dermst}) 
is a decreasing function of the density, the scalar
potential acting on the nucleons ($\propto M_{N0}-M^*_{N}$) will grow less rapidly than 
$\propto \rho$.
This ``decoupling mechanism'' of the scalar and vector potentials for higher densities is a
necessary condition for saturation in any Hartree-type theory \cite{WAL,GUI}. 

To appreciate this point, let us discuss the behavior of the $\sigma$ meson
mass as a function of the scalar mean field in more detail. As we discussed in sect. 1, the
Mexican hat shape of the vacuum effective potential (${\cal E}_V$) implies that the vacuum
fluctuations contribute attractively to the sigma mass. This can be seen from (\ref{msigma}), which
decreases with decreasing $M$ and becomes negative below some value of $M$ (tachyon pole). This
leads to an attractive contribution to the interaction
(\ref{f0}), increasing with density. The density dependent nucleon loop
contribution to the $\sigma$ mass in Eq.(\ref{second}) consists of two terms. The first one is 
the Fermi
average over the ``Z-graph'' contribution discussed in sect.1. It is repulsive and for 
the case of elementary
nucleons, where $g_{\sigma }$ is a constant independent of the scalar field, increases
rapidly with increasing scalar potential (decreasing $M_N$). Nevertheless, from $\sigma$ and NJL model 
studies for elementary fermions \cite{UNP}-\cite{BUB}
it is known that, irrespective of the model parameters, the size of this term is not sufficient 
to stabilize the state at normal densities - although for higher densities it does tend 
to stabilize the abnormal state.

The second term in (\ref{second}) is the Fermi average over a $\sigma \sigma $NN contact-type 
interaction,
which arises from the ``scalar polarizability'' ($\propto \frac{\partial^2 M_N}{\partial M^2}$) 
of the
nucleon - i.e., the response of the internal structure of the nucleon to the applied scalar
field. The presence of such a term, which vanishes in the case of elementary fermions, 
has been pointed out in a series of papers
\cite{ZGR,WALL}. If the curvature of the function 
$M_N(M)$ is positive everywhere and increasing with decreasing $M$, we can
expect that this term gives a stabilizing contribution which might lead to a stable normal state.
In this connection it should be emphasized that confinement effects are expected to lead to a nucleon mass
which does not vanish as the quark mass $M$ vanishes. That is, confinement effects should make the curve
$M_N(M)$ ``flatter'' than the naive case $M_N=3M$, in particular for small $M$. This would imply a positive curvature 
which increases, and
a slope which decreases, with increasing scalar potential - i.e., a repulsive
contribution due to the contact term and a suppression of the ``Z-graph''. This
important point will be investigated further in a simple model calculation in the next
section.

\subsection{Chiral symmetry and connection to the linear sigma model}
The contributions to the $\sigma$ mass in the medium can be expressed in terms of the
derivative of the quark condensate with respect to $M$, see Eq.(\ref{rel2}). For the
pion, the analogous relation is obtained from the low energy theorem 
$\langle \rho|{\overline \psi}
\psi|\rho\rangle = M\,\Pi_{\pi}(k=0)$, which holds in the NJL model between the quark
condensate and the pionic polarization operator at zero momentum (see e.g; Appendix B 
of Ref.\cite{LIGHT}).
Together with Eq.(\ref{dcond}), this shows that the pion mass in the medium, 
$M_{\pi}^2 \equiv g^2\left(\frac{1}{2G_{\pi}} + \Pi_{\pi}(k=0)\right)$, is related to the
derivative of the energy density as
\be
\frac{\partial {\cal E}}{\partial M}=\frac{M\,M_{\pi}^2}{g^2} - \frac{m}{2G_{\pi}}.
\label{pim}
\ee
At the physical point $M=M^*$ and for exact chiral symmetry ($m=0$), one therefore has the
possibilities $M_{\pi}^*=0$ (Goldstone mode) or $M^*=0$ (Wigner mode). Explicitly we have
from eqs.(\ref{hybc})-(\ref{cond}):
\begin{eqnarray}
M_{\pi}^2 &=& g^2 \left(\frac{1}{2G_{\pi}} -2i\gamma_q \kint{k}\frac{1}{k^2-M^2+i\epsilon}\right)
\label{pivac} \\
&+& \left(g_{\pi}^2 + g_{\pi \pi }\right) \gamma_N \int \frac{d^3k}{\left(2\pi\right)^3} 
n(k) \frac{1}{E_N(k)},
\label{pivde}
\end{eqnarray}
where (\ref{pivac}) is the vacuum term ($M_{\pi V}^2$), which leads to a tachyonic pion 
for $M<M_0$,  
and for the Fermi sea contributions we introduced the $\pi NN$ and $\pi \pi NN$ coupling constants
\begin{eqnarray}
\frac{g_{\pi }}{g} &=& \frac{M_N}{M} \label{pin1} \\
\frac{g_{\pi \pi }}{g^2} &=& \frac{M_N}{M} \left(\frac{\partial M_N}{\partial M} - 
\frac{M_N}{M} \right). \label{pin2}
\end{eqnarray}
The contribution proportional to $g_{\pi }^2$ in (\ref{pivde}) is the Fermi average over the 
``Z-graph'' contribution with external
zero momentum pion lines\footnote{This follows from the fact that the chiral Ward identity
implies ${\bold \Gamma}_{\pi }(0)=\frac{M_N}{M}\,g\,\gamma_5 {\bold \tau}$ for the $\pi NN$
vertex at zero pion momentum.
We also note that the 
coupling constant $g_{\pi }$ is {\em not} the physical $\pi NN$ coupling constant, but
it corresponds to the coupling constant appearing in the $\pi$ NN and $\sigma$ NN interaction
Lagrangians of the linear $\sigma$ model 
for elementary nucleons \cite{BLA}. In the linear $\sigma$ model for nucleons, the 
coupling constant appearing in the Lagrangian (denoted as $g_{\pi}$ here) is {\em defined} \cite{MR} 
by the relation $M_{N0}=g_{\pi } \, v$, where $v$ is the vacuum expectation value of the
renormalized $\sigma$ field, which in our notation is given by $v=M/g$.}, 
and the term proportional to $g_{\pi \pi}$, which vanishes in the ``naive'' case
$M_N\rightarrow 3M$, is the Fermi average over the $\pi \pi$NN contact term. 
It should be noted that the pionic ``Z-graph'' is {\em not} suppressed (but rather enhanced) 
by the mechanism discussed in the previous subsection, but nevertheless the sum of the Z-graph 
and the
contact term, which is proportional to $\frac{M_N}{M}\frac{\partial M_N}{\partial M}$, 
might be suppressed 
if the curve $M_N(M)$ becomes ``flat'' for small $M$.     

Using the expressions (\ref{rel2}) and (\ref{pim}) for the derivatives of the energy density
in terms of the sigma and pion masses, it is easy to show that that vacuum part, ${\cal E}_V$,
in Eq.(\ref{env}) is essentially equivalent to the ``Mexican hat'' of the
linear sigma model. Using (\ref{env}) we can write
${\cal E}_V(M)={\cal E}_1(M^2) - (M-M_0)\frac{m}{2 G_{\pi}}$, where ${\cal E}_1$ depends
only on $M^2$. Expanding ${\cal E}_1$ around $M^2=M_0^2$ we obtain
\begin{eqnarray}
{\cal E}_V(M)&=& - \left(M-M_0\right) \frac{m}{2 G_{\pi}} + 
\left(M^2-M_0^2\right) \left(\frac{\partial {\cal E}_1}{\partial M^2}\right)_{M^2=M_0^2}
\nonumber \\
&+& \frac{1}{2} \left(M^2-M_0^2\right)^2 \left(\frac{\partial^2 {\cal E}_1}{\partial \left(M^2\right)^2}
\right)_{M^2=M_0^2}+ {\cal O}\left(\left(M^2-M_0^2\right)^3\right). \nonumber \\
\label{ref}
\end{eqnarray}
Since the derivatives of ${\cal E}_1$ at $M=M_0$ are expressed as 
${\displaystyle \left(\frac{\partial {\cal E}_1}{\partial M}\right)_{M=M_0} = \frac{M\,M_{\pi 0}^2}{g^2}}$ and
${\displaystyle \left(\frac{\partial^2 {\cal E}_1}{\partial M^2}\right)_{M=M_0} = \frac{M_{\sigma 0}^2}{g^2}}$,
the above expression becomes
\begin{eqnarray}
{\cal E}_V(M) &=& - \left(M-M_0\right) \frac{m}{2 G_{\pi}} + \left(M^2-M_0^2\right)^2 
\frac{M_{\sigma 0}^2-M_{\pi 0}^2}{8M_0^2 g^2} + \left(M^2-M_0^2\right) \frac{M_{\pi 0}^2}{2g^2}
\nonumber \\
&+& {\cal O}\left(\left(M^2-M_0^2\right)^3\right).
\label{hat}
\end{eqnarray}
Except for the terms of third and higher order in $\left(M^2-M_0^2\right)$, this is the 
Mexican hat of the linear sigma model. 

\section{Numerical results and discussions}
\setcounter{equation}{0}
Let us now discuss the numerical results based on the expression (\ref{en1}) for
the energy density. First we refer to a calculation where confinement effects (removal
of unphysical thresholds) are not incorporated. The parameters of the model are the four-fermi 
coupling
constants $G_{\pi}$, $G_{s}$ and $G_{\omega}$, the current quark mass $m$, and the ultra
violet (UV) cut-off
$\Lambda$. $G_{\pi}$, $m$ and $\Lambda$ are determined as usual \cite{FAD1} by the conditions 
$m_{\pi}$=140 MeV, $f_{\pi}$=93 MeV, and $M_0$=400 MeV. 
(Here $m_{\pi}$ and $f_{\pi}$ are defined at the pion pole, in contrast to $M_{\pi 0}$ and $F_{\pi}$ used
earlier. We also note that to fix the solution of the gap equation at zero density as 
$M_0$=400 MeV is rather arbitrary, 
but qualitatively similar results are obtained for other
reasonable choices.) Lastly, $G_{s}$ (or the ratio $r_s=G_{s}/G_{\pi}$) is determined by 
requiring $M_{N0}$=939 MeV for the pole position of the quark-diquark t-matrix in the
static approximation (\ref{tn}) 
at zero density. In particular, for all cases considered here this requirement implies
a pole of the diquark t-matrix (\ref{taus}) at the scalar diquark mass $M_D$, and the
numerical calculation shows that this 
pole term, together with the elementary 4-fermi interaction $(4iG_s)$, overwhelms the
contribution due to the qq continuum. We will therefore, for simplicity, approximate 
the qq t-matrix as $\tau_s(q) \simeq 4iG_s + g_s/(q^2-M_D^2)$ in the following
calculations, where $g_s$ is the residue of (\ref{taus}) at the pole \cite{STAT2}.  
The parameters determined in this way for the Euclidean sharp cut-off, the 
3-momentum sharp cut-off, and the proper-time cut-off schemes are listed in Table 1.
$G_{\omega}$ is the only parameter which is left for the finite density calculation, and for all
cases considered here we
fix it by the requirement that the curve for the binding energy per nucleon ($E_B/A$) passes through 
the point $(\rho, E_B/A)=(0.16 fm^{-3}, -15 MeV)$. We first discuss the case of the
Euclidean sharp cut-off, since for this case the exact Faddeev results for the nucleon
mass are available \cite{FAD1} for comparison. 

Fig.1 shows the diquark mass (dot-dashed line), the quark-diquark threshold $M_D+M$ 
(upper double-dot dashed line), 
and the nucleon mass in the static approximation (lower dashed line) as functions of the 
scalar potential $\Phi = M_0-M$. (The dotted line shows $M=M_0-\Phi$ for orientation.) For comparison we also
show the result of the exact Faddeev calculation for the nucleon mass \cite{FAD1,PRIV} by the solid line. 
(In the exact Faddeev
calculation, a slightly larger ratio, $r_s=0.655$, has to be used in order to get the same 
nucleon mass at $\Phi=0$.) 

In all cases shown here, the nucleon pole exists only up to some
value of the scalar potential.  
Fig.1 shows that, in comparison to the exact Faddeev calculation, the result of the static approximation 
decreases much too fast with increasing scalar potential. 
In particular, since the quark-diquark interaction in the static approximation is
proportional to $1/M$, it develops an unphysical singularity as
$M\rightarrow 0$. The upper dashed line in Fig.1 shows the result if we fix the strength of the quark-diquark
interaction as $1/M_0$. This agrees much better with the exact Faddeev result for small $\Phi$, 
but for larger $\Phi$ the attraction is underestimated. Since the main purpose of this paper is to present a more qualitative
rather than quantitative discussion of the mechanism which leads to saturation in the NJL model, we will
use a simple interpolation between these two extreme cases ($1/M$ and $1/M_0$) for the quark-diquark interaction,
which avoids the unphysical singularity as $M\rightarrow 0$ and reproduces the exact Faddeev results for the
case of the Euclidean sharp cut-off quite well\footnote{One could also introduce a momentum dependence into
the quark-diquark interaction by some kind of form factor. However, also in this case one has 
to fit the parameters of the form factor to the exact Faddeev result, leading to a similar
situation as for our interpolated quark-diquark interaction strength.}. 
The middle dashed line in Fig.1 shows the
result obtained by using the function ${\displaystyle \frac{1}{M_0} \frac{M_0+c}{M+c}}$ with $c=700 MeV$ to
interpolate between the two extreme cases. In the following we will frequently refer 
to this interpolation (denoted case (ii)), but the cases
where the quark-diquark interaction is given by $1/M_0$ (case (i)) or  
$1/M$ (case (iii)) will also be discussed.  

In Fig.2 we show the resulting binding energy per nucleon as a function of the density 
for the three cases
(i), (ii), (iii) discussed above. Irrespective of the choice for $G_{\omega}$ there is no stable state
for cases (i) and (ii). The behavior of these curves is actually very similar to the one 
for the case
of elementary nucleons \cite{BLA} in the sense that their curvature is negative and  
there is no saturation in the range of densities where solutions
exist. For case (iii) some kind of saturation occurs, but it involves an unreasonably small nucleon mass
(see Fig.3) and large binding energy and density. This case resembles more the situation of the abnormal
state found for elementary nucleons \cite{BLA}. As is clear from the discussion in sect.2.3, a negative 
curvature of the function 
$M_N(M)$, such as that shown in Fig.1 for case (iii), corresponds to an attractive contact interaction 
in addition to the attractive vacuum contributions, and this reinforces the trend towards instability of
the normal state and a consequent transition to the abnormal state. Moreover, this case 
shows a pathological behavior for small densities which is caused by the fact that with this
choice of parameters (large $G_{\omega}$) the zero density value of $f_{0,L=0}$ is positive 
(i.e., dominated by the
$\omega$ meson exchange term), see Eq.(\ref{lowd}).  

Figure 3 shows the effective quark mass $M^*$ (dashed lines) and nucleon mass $M_N^*$ (solid lines) corresponding
to the three cases (i), (ii) and (iii), and in Fig.4 we plot the ``effective potential'' 
(${\cal E}(M)-M_{N0}\rho$)
as a function of $M$ for the cases (ii) and (iii) for three values of the density. 
For case (ii) there is no
minimum in the effective potential for densities beyond $\simeq 0.22 fm^{-3}$ in the range 
of $M$ values
where solutions for the nucleon mass exist, and in case (iii) the minimum rapidly 
moves towards small $M$ with increasing density. These curves show the de-stabilizing
effect of the attractive vacuum contributions (\ref{msigma}) to the effective $\sigma$ mass
: For larger scalar potential (smaller $M$)
the curvature of the lines in Fig.3 decreases, the minima become flatter and eventually
disappear.    

As we will see later, essentially the same situation is found with the proper-time cut-off scheme, and Appendix A
shows a similar situation for the sharp 3-momentum cut-off scheme.
For the case of elementary nucleons, it has been pointed out in several papers \cite{BOG,KOCH} that saturation
can be achieved by including higher order interaction terms, which eventually make the 
coupling constants appearing in the original model density dependent, leading to repulsive effects.
Actually, in Appendix A we show that in the present NJL model calculation the use of an 
8-fermi interaction of the form
\be
{\cal L}_8 = G_8 \left[\left({\overline \psi} \psi \right)^2 - 
\left({\overline \psi} \gamma_5 {\bold \tau} \psi 
\right)^2 \right] \left({\overline \psi} \gamma^{\mu} \psi \right)^2
\label{eight}
\ee
at the mean field level effectively leads to a renormalized vector coupling 
${\tilde G_{\omega}}$ which
increases with increasing scalar potential such that saturation can be achieved. Without sound
guiding principles, however, the introduction of higher order interaction terms is 
somewhat unsatisfactory
and will not be further discussed here\footnote{We have also checked 
that the problem of saturation cannot be solved by 
introducing a kind
of glueball field such as to make the NJL Lagrangian scale invariant \cite{SCALE}, nor by 
introducing the t'Hooft 6-fermi (determinant) interaction due to a nonzero strange quark 
condensate \cite{DET}.}.

Qualitatively the same instability has already been observed in the case of the 
$\sigma$ or NJL model 
for elementary fermions \cite{UNP}-\cite{BUB}. Actually, we can expect
from Fig.1 that we cannot improve the situation in the present NJL model calculation
for nucleons with internal quark structure: 
As we have pointed out in
sect.2.3, a qualitative difference (stabilization of the system) compared to the case of 
elementary nucleons could emerge if the curve $M_N(M)$ 
became flat for larger scalar potentials. However, Fig.1 demonstrates that as long as there exists 
a quark-diquark
threshold (or a quark-quark threshold at $3M$ if the diquark is unbound), there is only 
little room for this
effect to work\footnote{It is possible to find cases where $M_N(M)$ has a positive curvature 
(e.g; the case of the 
3-momentum sharp cut-off with $M_0\simeq 350 MeV$), but not large enough to lead to stabilization. 
Moreover, in these
cases the binding energy of the nucleon decreases and its size becomes larger and larger 
with increasing scalar potential, which is also physically unacceptable.}.

The presence of unphysical thresholds is a consequence of the lack
of confinement in the NJL model. In particular, the thresholds force the nucleon mass to
approach zero as the constituent quark mass goes to zero. In contrast to this situation,
in models with confinement (bag models etc.) the nucleon mass is nonzero even (if the mass
of the constituents is zero) because of the finite kinetic energy of the quarks inside the cavity.
From this viewpoint, the presence of an appreciably positive scalar polarizability of the
nucleon at larger scalar potentials can be considered as a manifestation of confinement.  
 
We will therefore consider a simple method to avoid unphysical thresholds, which was proposed 
in Ref.\cite{INFR}.
This consists of introducing an infrared (IR) cut-off ($\mu$) in addition to the UV cut-off ($\Lambda$) in the
proper time regularization scheme. (Some formulae are collected in Appendix B, where it is
also shown explicitly that the thresholds of the quark-quark and quark-diquark loops are removed
by the IR cut-off.) In the following discussion of the numerical results, we will concentrate
on the comparison between the cases $\mu=0$, for which we will observe essentially the same
situation as for the Euclidean cut-off scheme discussed above, and $\mu>0$. Therefore we
confine ourselves to the case (ii) discussed above - i.e., the case where the strength of the
quark-diquark interaction to be used in the static approximation is adjusted to agree
quite well with the exact Faddeev result in the Euclidean cut-off scheme\footnote{Unfortunately, 
exact Faddeev results for the proper time regularization scheme are not yet available. We should also note that 
the IR cut-off removes the unphysical thresholds only in the proper time regularization scheme. Besides this, the proper time 
scheme has certain advantages, such as preserving gauge invariance, but it has the disadvantage that in general 
it leads to poles of Green functions in unphysical regions of the complex plane, 
which can cause difficulties - 
e.g; in the calculation of form factors.}. 
Very similar results
are found for case (i), but not for case (iii) which suffers from the unphysical singularity of
the quark-diquark interaction as discussed above. We will fix the value of the
IR cut-off as $\mu=200 MeV$. (The results are very similar as long as $\mu > 100 MeV$.)
Once $\mu$ is fixed, the other parameters are determined in the same way as discussed above,
and their values are listed in Table 1.   
    
Fig.5 shows the plot for $M_N(M)$ for the two cases $\mu=0$ (lower solid curve) and $\mu=200 MeV$
(upper solid curve). Also shown are the diquark mass $M_D(M)$ and the sum $M_D(M)+M$ for the two cases.
It is interesting to observe that, although the IR cut-off also cancels the threshold 
in the quark-quark loop, the behavior of the diquark mass $M_D(M)$ is almost unchanged by the
IR cut-off (see Appendix B), but the behavior of the nucleon mass $M_N(M)$ is changed drastically.
One could interpret this by saying that the confining mechanism is not at work for the
quarks in the diquark but only for the system as a whole, which is physically reasonable.
The case $\mu=200 MeV$ shows an appreciable scalar polarizability of the nucleon for larger
scalar potentials, and 
accordingly the nucleon mass does not tend to zero with decreasing constituent quark mass.
As we have discussed in detail in sect.2.3, this implies a repulsive contribution to the sigma
meson mass and a reduction of the $\sigma NN$ coupling in the medium, working towards stabilization of the system.
  
In Fig.6 we show the binding energy per nucleon for the same two cases: $\mu=0$ (dashed line)
and $\mu=200 MeV$ (solid line). The corresponding values of $G_{\omega}$ (see Table 1) have 
been adjusted such that the curves
pass through the point $(\rho,E_B/A)$ = $(0.16 fm^{-3}, -15 MeV)$. The result for $\mu=0$ (dashed line) 
is very similar to the one obtained with
the Euclidean sharp cut-off (the line labelled case (ii) in Fig.3) or the 3-momentum cut-off
(Appendix A). The case $\mu=200 MeV$, on the
other hand, leads to an equation of state which saturates. The fact that the calculation does not
reproduce the empirical saturation point (the saturation density is too high) is related to the
fact that we have only one free parameter ($G_{\omega}$) in the finite density calculation.
Nevertheless, the observation that the inclusion of confinement aspects leads to a nucleon mass
which has positive curvature (scalar polarizability) as a function of the scalar potential, 
which in turn leads to a saturating binding energy, indicates that the
long standing problem of matter stability in chiral models with an effective vacuum potential of
the Mexican hat shape 
can be solved by taking into account the quark structure of the nucleon. 

The effective masses of the quark and the nucleon are shown in Fig.7 for the same two cases
($\mu=0$: dashed lines, $\mu=200 MeV$: solid lines). As we will discuss in more detail later, 
their behavior
can be understood from the relations (\ref{dermqst}) and (\ref{dermst}), in terms of the density
dependence of the quark-nucleon and nucleon-nucleon interaction. For the case $\mu=0$, the
attraction due to the $\sigma$ meson exchange increases with increasing density due to the
tachyon pole in the $\sigma$ propagator, while for the
case $\mu=200 MeV$ it decreases due to the repulsive $\sigma NN$ contact term and the reduced
$\sigma NN$ coupling, similarly to the case of phenomenologically successful
non-chiral models \cite{WAL,GUI}.  
Fig.8 shows the ``effective potential'', ${\cal E}(M)-M_{N0}\rho$, for the two cases as  
functions of $M$ for three values of the density. We see that for any fixed density
the curvature of the effective potential decreases much faster with increasing scalar
potential for the case $\mu=0$ than for the case $\mu>0$. This demonstrates again the 
important stabilizing role played by the $\sigma NN$ contact term. 

Let us now analyze the difference between the cases $\mu=0$ and $\mu>0$ in more detail.
In Fig.9 we show the density dependent contributions to the squared $\sigma$
mass (Eq.(\ref{second})) for $\rho=0.16 fm^{-3}$ as functions of $M$. The solid lines
show the ``Z-graph'' contributions multiplied by a factor 10, and the dashed lines refer
to the contact terms. By comparing the two solid lines we clearly see the suppression
of the Z-graph contribution caused by the reduced slope of $M_N(M)$ (reduced coupling 
constant $g_{\sigma }$) for the case $\mu>0$. The largest and most important contribution,
however, is due to the contact term in the case $\mu>0$, which increases with increasing
scalar potential due to the increasing curvature of $M_N(M)$. The role
of this term is further illustrated in Fig.10, which shows the $\sigma$ propagator
in the medium at zero momentum ($\left(\frac{\partial^2 {\cal E}}{\partial M^2}\right)^{-1}$, 
dashed
lines) and the Landau-Migdal parameter $f_{L=0}$ (Eq.(\ref{f0}), solid lines),  
for $\rho=0.16 fm^{-3}$ as functions of $M$. Since for the case $\mu=0$ the density
dependent contributions to the $\sigma$ mass are rather small, the vacuum term plays the
dominant role, i.e., there is a
tachyon pole in the $\sigma$ propagator, and the attraction due to $\sigma$
meson exchange grows without limits with increasing scalar potential, causing the
collapse of the system. In the case $\mu>0$, however, the tachyon
pole is avoided because of the large repulsive contribution associated with the 
contact term, leading to
a smooth decrease of the $\sigma$ mass with increasing scalar potential. Moreover, due to 
the reduction of the $\sigma NN$ coupling constant, the $\sigma$ meson exchange contribution
to the Landau-Migdal parameter $f_{L=0}$ now decreases smoothly with increasing scalar potential,
similarly to the case of non-chiral models. 

Fig.11 shows the same quantities at $M=M^*(\rho)$ as functions
of the density. For $\mu>0$ the $\sigma$ mass in the medium is almost constant, and
$f^*_{L=0}$ changes smoothly from attraction at small densities (dominance of the $\sigma$
meson exchange term in Eq.(\ref{f0})) to repulsion at larger densities (dominance of the $\omega$
meson exchange term).  
It is important to realize that the density dependence of $f^*_{L=0}$ is directly related to the saturation of the
binding energy per nucleon ($E_B/A={\cal E}^*/\rho-M_{N0}$). Since for small densities 
the curvature of $E_B/A$ has to be negative (see Eq.(\ref{lowd})), the necessary condition
for saturation is that the curvature of $E_B/A$ turns from negative to positive
at some density $\rho_c$, which is the well-know liquid-gas phase transition. 
From this it is easy to show\footnote{To see this, one expands both side of the 
relation $\rho (E_B/A)''=-2(E_B/A)' + {\cal E}''$
around $\rho=\rho_c$ up to first order in $(\rho-\rho_c)$, and uses ${\cal E}''=\pi^2 v^*_F/2 p_F^2 + f^*_{L=0}$
as well as the fact that at $\rho=\rho_c$ the function $(E_B/A)'$ has a minimum. (Here the primes indicate
derivatives w.r.t. the density.)}
that the necessary conditions for saturation can be
expressed as $f^*_{L=0}(\rho_c)<0$ and $\left(\frac{{\rm d}f^*_{L=0}}{{\rm d}\rho}\right)_{\rho=\rho_c}
>0$. It is clear from Fig.11 that the case $\mu=0$ does not satisfy this
condition.
      
Finally, in Fig.12 we plot the one-body $\sigma$-term, $\Sigma^*_{\pi N}$, 
and the two-body $\sigma$-term, 
$\Sigma^*_{\pi NN}$, defined in (\ref{sigma12}) by the solid and dashed lines, respectively,
and also the sum $\frac{\pi^2}{2 p_F^2} 
\Sigma^{'*}_{\pi N}+\Sigma^*_{\pi NN}$ which appears in eqs. (\ref{exdco})-(\ref{exdmn}) by 
the dot-dashed lines 
as functions of the density for the cases $\mu=0$ and $\mu>0$. Referring to Eq.(\ref{onesig}),
we understand that the decrease of 
$\Sigma^*_{\pi N}$ with increasing density for the case $\mu>0$ is caused by the
decrease of the $\sigma NN$ coupling constant, while the sharp increase for the case
$\mu=0$ reflects the increasing attraction due to the decreasing $\sigma$ mass (tachyon pole). The
zero density values in both cases are $\Sigma_{\pi N,0} \simeq 2.7 \times m \times
0.7 \simeq 32 MeV$, where the factors 2.7 and 0.7 correspond to  
$\left(\frac{\partial M_N}{\partial M}\right)_{M=M_0}$ and $\frac{{\rm d}M_0}{{\rm d}m}$ in Eq.(\ref{onesig}), 
respectively.
The fact that this is too small compared with the experimental value  
$\Sigma_{\pi N,0} = 45 \pm 5 MeV$
is partially related to the fact that in the presently used proper-time 
regularization scheme we are in the
``firmly broken'' regime \cite{FIRM} where $\frac{{\rm d}M_0}{{\rm d}m}<1$, and partially
to the fact that we did not include the effect of the pion cloud \cite{CLOUD}. 
The sign of the two-body $\sigma$ term ($m {\rm d}f^*_{L=0}/{\rm d} m$)
is determined by the sign of $\left(\partial f_{L=0}/\partial M\right)_{M=M^*}$. For small
densities, $f_{L=0}(M)$ exhibits the singularity due to the tachyon pole; i.e., it decreases
rapidly with decreasing $M$. For the case $\mu=0$, this behavior also persists for
larger densities (see Fig.10), and therefore $\left(\partial f_{L=0}/\partial M\right)_{M=M^*}$
is positive, increasing rapidly with increasing density. For the case $\mu>0$, however,
the large contact term removes the tachionic behaviour (see Fig.10), and 
$\left(\partial f_{L=0}/\partial M\right)_{M=M^*}$ becomes a smooth function changing sign at 
$\rho\simeq 0.07 fm^{-3}$. This is reflected by the behavior of $\Sigma^*_{\pi NN}$ shown
in Fig.12.    

\section{Summary and conclusions}
\setcounter{equation}{0}
In this paper we investigated whether the old problem of matter stability in chiral
field theory models, based on a linear realization of the symmetry, can be solved
by taking into account the quark structure of the nucleon. The basic problem in
these models is associated with the Mexican hat shape of the vacuum effective potential,
which implies strongly attractive contributions to the $\sigma$ meson mass from
vacuum fluctuation effects. This decrease of the $\sigma$ meson mass with increasing
scalar potential leads to an attractive Landau-Migdal interaction between the nucleons 
which increases with increasing density. Since the necessary condition for saturation 
in any relativistic mean field theory is that the attraction should decrease at 
high densities (decoupling of the $\sigma$ and $\omega$ mesons), it follows that
stable normal matter cannot be described. If the nucleons are treated as elementary
fields, it is possible to stabilize the matter by taking into account effects involving 
the polarization of the Dirac sea of the nucleon, but these give extremely large
contributions which change the overall physical picture completely, throwing doubts
on the reliability of this treatment.

  To incorporate the structure of the nucleon into the calculation of the nuclear
matter equation of state, we used the Nambu-Jona-Lasinio (NJL) model to describe the
nucleon as a quark-diquark state. We then employed a hybrid description
of nuclear matter, in which the nucleons are moving in self consistent scalar
and vector fields which couple to the quarks, and the polarization of the Dirac 
sea of quarks was taken into account. By considering various expansions in powers
of the density, as well as the Landau-Migdal interaction between the nucleons,
we have demonstrated that this description is very similar to that of the
linear $\sigma$ model for elementary nucleons. In fact, there is essentially only one
important difference. If the scalar field acts on the quarks instead of
directly on the nucleons, the nucleon mass need not be a linearly decreasing
function of the scalar potential - i.e., there can be a curvature, which is equivalent to
a ``scalar polarizability'' of the nucleon. The presence of a positive curvature of the
nucleon mass as a function of the scalar potential is also suggested by confinement.
The constituent quark mass decreases linearly with the scalar potential, but if the 
quarks are confined in the nucleon the mass of the nucleon will not decrease towards zero
simultaneously with the quark mass. If such a scalar polarizability of the nucleon
exists, it has two important consequences. First, there will be a reduction of the $\sigma NN$ 
coupling constant, and second a repulsive contribution to the $\sigma$ meson mass.
Both effects will increase with increasing scalar potential (density), and work against the
attractive vacuum fluctuation effects discussed above - i.e., towards the stabilization
of the system. 

One of the shortcomings of usual treatments based on the NJL model is, however,
the absence of confinement, leading to unphysical thresholds for bound states like
the nucleon. We have seen that, as long as these thresholds are present, little
room is left for the effect discussed above, and one encounters essentially the same
situation as in the elementary nucleon case. The Landau-Migdal interaction becomes
more and more attractive as the density increases, causing the collapse of the
system. Therefore, in order that the stabilizing effect of the scalar polarizability
can come into play, one has to avoid the unphysical thresholds. Several methods
have been proposed earlier for this purpose, and for the qualitative purpose of this
work we have chosen the most simple one, which consists in introducing an infrared
cut-off in addition to the ultraviolet cut-off, in the framework of the proper-time 
regularization scheme. We have shown that for physically reasonable values of this
infrared cut-off the nucleon mass as a function of the scalar potential does indeed have
an appreciable curvature, in particular for large scalar potentials. We have demonstrated
that this effect can lead to a saturating nuclear matter equation of state. That is, we
have confirmed that effects based on the quark structure of the nucleon work
towards a solution of the long standing problem of matter stability in chiral models
based on the linear realization of the symmetry. 

Before drawing firm conclusions,
however, it should be noted that for technical reasons our treatment is still
afflicted with an ambiguity. That is, in order to simplify the Faddeev treatment of the
quark-diquark state in the scalar potential, we used the static approximation to the
Faddeev equation, where the momentum dependence of the quark exchange kernel is
neglected. While this approximation works reasonably well in free space, in the
medium it becomes worse due to the reduced constituent quark mass. Therefore we 
used an interpolated strength for the static quark exchange kernel, adjusted so
as to reproduce the exact Faddeev result, which at present is available only for the
Euclidean sharp cut-off scheme. We then used this interpolated form
of the quark exchange kernel for other cut-off schemes as well. To resolve this
ambiguity, one should either use the exact Faddeev equation or introduce an approximation
which is valid for small constituent quark masses. Nevertheless, the results
of this paper strongly suggest that in a more complete treatment one should be able to
describe stable nuclear matter in a relativistic mean field treatment of chiral
theories by letting the mean field couple to the quarks, provided that a kind of
confinement mechanism ensures that the nucleon mass does not tend to zero with
increasing scalar potential.

\vspace{1 cm}

{\sc Acknowledgement}\\
This research was supported by the Australian Research Council, Adelaide University, 
and the Japanese Society for the Promotion of Science.
The authors wish to thank N. Ishii for providing them with the exact Faddeev results shown in 
Fig.1.
They also wish to thank P. A. M. Guichon, K. Saito, K. Tsushima and K. Yazaki for helpful discussions. 
One of the authors (W.B.) thanks M. Birse for an interesting discussion on Z-graphs, and 
O. Morimatsu for communications on QCD sum rules.

\newpage
\appendix
{\LARGE Appendices}

\section{Results for the 3-momentum sharp cut-off scheme and stabilization 
due to higher order interaction terms}
\setcounter{equation}{0}
In this appendix we show the results obtained in the 3-momentum sharp cut-off
scheme, including the stabilizing effect due to the 8-fermi interaction term
(\ref{eight}). (The stabilization due to the 8-fermi interaction is, however,
independent of the regularization scheme.)

Defining for the purpose of this appendix the field 
$\sigma\equiv \langle \rho|{\overline \psi} \psi|\rho\rangle$ in addition to the
vector field $\omega^{\mu}= \langle \rho|{\overline \psi}\gamma^{\mu} \psi|\rho\rangle$,
the inclusion of the 8-fermi interaction (\ref{eight}) on the mean field level 
modifies the Lagrangian (\ref{lag1}) to
\be
{\cal L}={\overline \psi}\left(i \fslash{\partial}- M - 2 {\tilde G}_{\omega} \gamma^{\mu} \omega_{\mu} 
\right) \psi - G_{\pi} \sigma^2 + G_{\omega} \omega_{\mu}\omega^{\mu} -3 G_8 \sigma^2 \omega_{\mu}\omega^{\mu}
+ {\cal L}_I,
\label{lag11}
\ee
where $M=m-2{\tilde G}_{\pi} \sigma$, ${\tilde G}_{\omega}=
G_{\omega} - G_8 \sigma^2$ and ${\tilde G}_{\pi}=G_{\pi} + G_8 \omega_{\mu}\omega^{\mu}.$  
Since the nucleon energy has the form (\ref{spec}) with $G_{\omega}\rightarrow {\tilde G}_{\omega}$,
the expression for the energy density in the hybrid model becomes (cf. Eq.(\ref{en}))
\be
{\cal E}={\cal E}_{Vq}(M) + G_{\pi}\sigma^2 - G_{\omega} \omega_{\mu}\omega^{\mu} 
+3 G_8 \sigma^2 \omega_{\mu}\omega^{\mu} + 6{\tilde G}_{\omega} \omega^0 \rho
+{\cal E}_F,
\ee
where 
${\cal E}_{Vq}(M)$ is the quark loop term (the first term in Eq.(\ref{env})), and ${\cal E}_F$
is given by (\ref{enf}), where now ${\bold k}_N={\bold k}-6{\tilde G}_{\omega}{\bold \omega}$. 
To calculate the variation w.r.t. $\sigma$ and $\omega^{\mu}$, one has to take into account that
now $M$, which appears in the quark loop as well as the nucleon Fermi motion term due to $M_N=M_N(M)$, as well as 
${\bold k}_N$ depend on both the scalar and vector fields. We obtain
\begin{eqnarray}
\frac{\partial {\cal E}}{\partial \omega_{\mu}}&=&-4G_8\omega^{\mu}\sigma \frac{\partial {\cal E}}{\partial M}
-2G_{\omega} \omega^{\mu} +6 G_8 \sigma^2 \omega^{\mu} + 6{\tilde G}_{\omega} j_B^{\mu} = 0 \label{var1} \\ 
\frac{\partial {\cal E}}{\partial \sigma} &=& 2{\tilde G}_{\pi} \frac{\partial {\cal E}}{\partial M}
-2G_{\pi}\sigma - 6 G_8\sigma \omega_{\mu}\omega^{\mu} + 12 G_8 \sigma \omega_{\mu} j_B^{\mu}  = 0,
\label{var2}
\end{eqnarray}
where the baryon current is as defined below Eq.(\ref{spec}), and the derivative w.r.t. $M$ acts on
the quark loop and nucleon Fermi motion terms (${\cal E}_{Vq}$ and ${\cal E}_F$).  
One can now show that $\omega^{\mu}=3 j_B^{\mu}$
is still a solution. If we directly insert this relation into (\ref{var1}) and (\ref{var2}), 
both equations give the same requirement, $\sigma = \partial {\cal E}/\partial M$. Using $\omega^{\mu}=3 j_B^{\mu}$ 
to eliminate
the vector field, and setting ${\bold j}_B=0$ for nuclear matter at rest, we can write the expression 
for the energy density in the same form as Eq.(\ref{en1}) for ${\bold j}_B=0$, where $G_{\pi}$ in the
denominator of the second term in (\ref{env}) should be replaced 
by ${\tilde G}_{\pi}=G_{\pi}\left(1+\beta \rho^2\right)$, with $\beta=9 G_8/G_{\pi}$.  
To discuss the stabilizing effect of $G_8$, it is more instructive, however, to combine the shift
$\displaystyle{\frac{(M-m)^2}{4}\left(\frac{1}{{\tilde G}_{\pi}}-\frac{1}{G_{\pi}}\right)}$ with the the term
$9G_{\omega}\rho^2$ of (\ref{enw}). This leads to the final result that the energy density is given by the
same expression as before, but with ${\cal E}_{\omega}$ replaced by
\be
{\tilde {\cal E}}_{\omega}=9G_{\omega}\rho^2\left(1-\frac{(M-m)^2}{4 G_{\pi}^2} 
\frac{G_8/G_{\omega}}{1+\beta \rho^2}\right), \,\,\,\,\,\,\,\,\,\,(\beta=9 G_8/G_{\pi}). 
\label{rep}
\ee
If we impose the condition $\displaystyle{1-\frac{(M_0-m)^2}{4 G_{\pi}^2} \frac{G_8}{G_{\omega}}>0}$, then
the term in the brackets of (\ref{rep}) is always positive for $M<M_0$. (This condition corresponds to
${\tilde G}_{\omega}>0$.) The important point is that the repulsion (\ref{rep}) increases with
decreasing $M$, or in other words, with increasing scalar potential (density). If $G_8$ is
chosen large enough, this term will ensure that the curves in Fig.4 will go up on the
low $M$ side and the minimum in the effective potential will not disappear. The calculation
shows that this is the case for $\beta>1 fm^6$. 

In Fig.13 we show the nucleon and diquark masses as functions of the scalar potential for the
3-dimensional sharp cut-off scheme, using case (ii) for the interpolated quark-diquark
interaction strength as discussed in sect. 3. (The parameters are given in Table 1.) 
The behavior of the solid line in Fig.13 is
very similar to the one labeled by (ii) in Fig.1 and therefore we cannot obtain a stable
nuclear matter ground state for $G_8=0$, as shown by the dashed line in Fig.14. The solid
line shows the result for $\beta=1.5 fm^6$. We wee that the binding energy indeed saturates, but
at a too low density. (It turns out that this cannot be improved by choosing a different value for 
$\beta$.) In Fig.15 we show the effective quark and nucleon masses for the cases $\beta=0$ and
$\beta=1.5 fm^6$. The
stabilizing effect of the 8-fermi interaction term is clearly seen.

\section{Some explicit formulae for the proper time regularization scheme}
\setcounter{equation}{0}
To apply the proper time regularization scheme to a momentum loop integral \cite{INFR},
one first follows the usual procedure of covariant loop integration,
i.e., one combines the denominators by introducing Feynman parameters, introduces
a shift of the loop momentum and performs the Wick rotation. In the resulting
expressions one then makes the replacements
\begin{eqnarray}
{\rm ln} A &\rightarrow& - \int_{1/\Lambda^2}^{1/\mu^2} \frac{{\rm d}\tau}{\tau} e^{-\tau A}
\nonumber \\
\frac{1}{A^n} &\rightarrow& \frac{1}{(n-1)!}\int_{1/\Lambda^2}^{1/\mu^2} {\rm d}\tau \tau^{n-1}
e^{-\tau A} \,\,\,\,\,\,\,\,\,\,(n\geq 1) \nonumber,
\end{eqnarray}
where $A$ depends on the momenta and Feynman parameters, and $\Lambda$ and $\mu$ denote the
UV and IR cut-off, respectively. Applying this procedure to the quark loop term
in the energy density (first term in (\ref{env})), the scalar qq bubble graph (\ref{bubbs}) and the
quark-diquark bubble graph (\ref{bubbn}), we obtain the expressions
\begin{eqnarray}
{\cal E}_{Vq}&=&\frac{\gamma_q}{16 \pi^2} \left(C_3(M^2)-C_3(M_0^2)\right) \label{envq} \\
\Pi_s(q)&=&-\frac{\gamma_q}{16 \pi^2}\left(2 C_2(M^2) + q^2 C_1(M^2) - q^2 \int_0^1 {\rm d}x \,x
\left(1-2x\right)\right. \nonumber \\
&\times &\left. \frac{e^{-A(M^2,M^2)/\Lambda^2} - e^{-A(M^2,M^2)/\mu^2}}{A(M^2,M^2)}\right)
\label{pisr} \\
\Pi_N(q)&=&-\frac{M G_S}{4 \pi^2} C_2(M^2) - \frac{g_s^2}{16 \pi^2}\left[\left(
\frac{\fslash{q}}{2}+M\right)C_1(M_D^2) + \int_0^1 {\rm d}x\,x\left(\frac{x}{2}\fslash{q}+M\right) \right.
\nonumber \\
& \times& \left.\left(M_D^2-M^2-q^2\left(1-2x\right)\right) 
\frac{e^{-A(M^2,M_D^2)/\Lambda^2}-e^{-A(M^2,M_D^2)/\mu^2}}{A(M^2,M_D^2)}\right].
\label{pinr}
\end{eqnarray}
Here we defined $C_n(m^2)=\int_{1/\Lambda^2}^{1/\mu^2} \frac{{\rm d}\tau}{\tau^n} e^{-\tau m^2}$,
and the $q^2$ and $x$-dependent functions $A(m_1^2,m_2^2)=m_1^2+(m_2^2-m_1^2)x-q^2x(1-x))$.
The diquark t-matrix $\tau_s$ has been approximated by the constant + pole terms,
($\tau_s(k)\simeq 4iG_s + g_s/(k^2-M_D^2)$), as has been explained in sect. 3.

It is clear from the above expressions that the imaginary parts of $\Pi_s$ and $\Pi_N$ are canceled
if $\mu^2>0$. It is also easy to verify that $\Pi_s(q^2)<0$ is a monotonically decreasing (increasing)
function of $q^2$ $(M^2)$. Therefore, the condition that in (\ref{taus}) there is a diquark pole 
($M_D^2>0$), even for $M=0$,
is $\Pi_s(q^2=0,M^2=0)>-1/(2G_s)$, which in turn requires 
$\Lambda^2-\pi^2/(3G_s) < \mu^2 < \Lambda^2 - \pi^2/(3G_{\pi})$. (The second inequality must be satisfied
in order to have a non-trivial solution of the gap equation at zero density.) The parameters
used in sect. 3 do not satisfy the first inequality, which means that the aspect of 
``confinement'' (i.e.,
to have a finite mass of the composite system even if the mass of the constituents tends to zero) 
is not present for the diquark. It is, however, present for the nucleon as a whole, as is indicated by
the results shown in Fig.5.

\newpage

\begin{table}[h]
\begin{center}
\begin{tabular}{|c|c||c||c|c|}
\hline
                            &  Eucl.   & 3-mom. & \multicolumn{2}{c|}{proper time} \\
                            &          &        & $\mu=0$  &  $\mu=200$ MeV \\
\hline 
m[MeV]                      &  8.99    & 5.96   &  17.08   &  16.93 \\
$\Lambda$ [MeV]             & 739.0    & 592.7  &  636.7   &  638.5 \\
$G_{\pi}$ [GeV$^{-2}$]      & 10.42    & 6.92   &  19.76   &  19.60 \\
$r_s=G_s/G_{\pi}$           & 0.65     & 0.73   &  0.51    &  0.49  \\
\hline
\end{tabular}
\end{center}
\caption{Parameters used for the Euclidean sharp cut-off (``Eucl.'') the 3-momentum
sharp cut-off (``3-mom.'') and the proper time regularization schemes.
For the proper time regularization scheme, we give two parameter sets with different IR 
cut offs $\mu$.}
\end{table}

\newpage

\section*{Figure Captions}
\begin{enumerate}
\item The solid and dashed lines show the nucleon mass $M_N$ as functions of the
'scalar potential' $\Phi=M_0-M$ in the Euclidean sharp cut-off scheme. The solid line is the exact
Faddeev result ($r_s=0.655$), and the dashed lines refer to the static approximation
($r_s=0.646$, other parameter given in Table 1) for the three cases of the quark-diquark 
interaction strength discussed 
in the main text. Upper dashed line: case (i), middle dashed line: case (ii), lower
dashed line: case (iii). The dot-dashed line shows the diquark mass $M_D$
for the case $r_s=0.646$, and the double-dot dashed line shows the quark-diquark
threshold for this case. The dotted line shows the quark mass $M=M_0-\Phi$ for
orientation.
\item The binding energy per nucleon as function of the density in the Euclidean sharp cut-off scheme 
for the three cases of the quark-diquark interaction strength discussed in the main text.
For each case, $G_{\omega}$ has been adjusted such that the curve passes through the
point $(\rho, E_B/A)=(0.16 fm^{-3}, -15 MeV)$, leading to $9G_{\omega}/G_{\pi}$=
5.68, 6.91 and 11.97 for cases (i), (ii) and (iii), respectively.    
\item The effective quark mass (dashed lines) and nucleon mass (solid lines) as
functions of the density in the Euclidean sharp cut-off scheme for the three cases of the 
quark-diquark interaction strength discussed in the main text. Upper lines:
case (i), middle lines: case (ii), lower lines: case (iii). The values of $G_{\omega}$
used in these three cases are given in the caption to Fig.2.
\item The 'effective potential' ${\cal E}-M_{N0} \rho$ as function of $M$ for four values
of the density in the
Euclidean sharp cut-off scheme for the cases (ii) and (iii) of the quark-diquark interaction 
strength discussed in the main text. The values of $G_{\omega}$
used in these two cases are given in the capture to Fig.2. The values of the density are:
$\rho=0$ (dotted line), $\rho=0.08 fm^{-3}$ (dashed lines), $\rho=0.16 fm^{-3}$ (solid lines),
$\rho=0.24 fm^{-3}$ (double-dot dashed lines).
\item The solid lines show the nucleon mass $M_N$ as function of the 'scalar potential'
$\phi=M_0-M$ in the proper-time regularization scheme,
referring to case (ii) for the quark-diquark interaction strength.
Lower solid line: case (a) ($\mu=0$, $r_s=0.492$), upper solid line: case (b) ($\mu=200 MeV$, $r_s=0.508$). 
Also shown are the diquark mass $M_D$ and the sum $M_D+M$ for these two
cases. 
The dotted line shows the quark mass $M=M_0-\Phi$ for orientation.
\item Binding energy per nucleon as function of the density in the proper-time
regularization scheme, referring to case (ii) for the quark-diquark interaction strength.
Shown are the results for the following two cases: (a) $\mu=0$ ($r_s=0.492$, $9G_{\omega}/G_{\pi}$=
3.971, dashed line), (b) $\mu=200 MeV$ ($r_s=0.508$, $9G_{\omega}/G_{\pi}$=
3.323, solid line). 
\item Effective quark and nucleon masses as functions of the density for the cases
(a) (solid lines) and (b) (dashed lines) described in the caption to Fig.6.
\item The 'effective potential' ${\cal E}-M_{N0} \rho$ as function of $M$ for four values
of the density in the proper-time regularization scheme, referring to the cases (a) and
(b) described in the caption to Fig.6. The values of the density are as in Fig.4:
$\rho=0$ (dotted line), $\rho=0.08 fm^{-3}$ (dashed lines), $\rho=0.16 fm^{-3}$ (solid lines),
$\rho=0.24 fm^{-3}$ (dash-double dotted lines). (The curve for the case $\rho=0$
is practically the same for the cases (a) and (b), and the dotted line here refers to
case (b).)
\item The solid lines show the ``Z-graph'' contributions to 
$\frac{\partial^2 {\cal E}}{\partial M^2}$, multiplied by a factor 10, and the dashed 
lines show the contributions of the contact term to $\frac{\partial^2 {\cal E}}{\partial M^2}$
as functions of $M$ for the density $\rho=0.16 fm^{-3}$. The cases (a) and (b) refer to 
the two cases described in the caption to Fig.6.    
\item The dashed lines show $\left(\frac{\partial^2 {\cal E}}{\partial M^2}\right)^{-1}$
and the solid lines show $f_{L=0}$ as functions of $M$ for the density $\rho=0.16 fm^{-3}$. 
The cases (a) and (b) refer to the two cases described in the caption to Fig.6.
\item The dashed lines show $\left(\frac{\partial^2 {\cal E}}{\partial M^2}\right)_{M=M^*}^{-1}$,
and the solid lines show $f^*_{L=0}$ as functions of the density. 
The cases (a) and (b) refer to the two cases described in the caption to Fig.6.
\item The solid lines show $\Sigma^*_{\pi N}$, the dashed lines show
$\Sigma^*_{\pi NN}$, and the dash-dotted lines show
$\Sigma^*_{\pi NN}+\frac{\pi^2}{2 p_F^2}\Sigma^{*'}_{\pi N}$ as functions of the density. 
The cases (a) and (b) refer to the two cases described in the caption to Fig.6.
\item The solid line shows the nucleon mass $M_N$ as function of the
'scalar potential' $\Phi=M_0-M$ in the 3-momentum sharp cut-off scheme, employing the static
approximation to the quark exchange kernel and the strength of the quark-diquark interaction
as in 'case (ii)' described in sect. 3. 
The dashed line shows the diquark mass $M_D$, and the dot-dashed line shows the quark-diquark
threshold. The dotted line shows the quark mass $M=M_0-\Phi$ for
orientation.
\item The binding energy per nucleon as function of the density in the 3-momentum
sharp cut-off scheme, using the nucleon mass as shown by the solid line in Fig.13. 
The dashed line shows
the result for $G_8=0$, and the solid line refers to the case $9G_8/G_{\pi}=1.5 fm^6.$  
For each case, $G_{\omega}$ has been adjusted such that the curves passes through the
point $(\rho, E_B/A)=(0.16 fm^{-3}, -15 MeV)$, leading to $9G_{\omega}/G_{\pi}$=
9.84 and 22.01 for the cases $G_8=0$ and $9G_8/G_{\pi}=1.5 fm^6$, respectively.
\item The effective quark mass (dashed lines) and nucleon mass (solid lines) as
functions of the density in the 3-momentum sharp cut-off scheme for the 
cases $G_8=0$ and $9G_8/G_{\pi}=1.5 fm^6$. 
    
\end{enumerate}

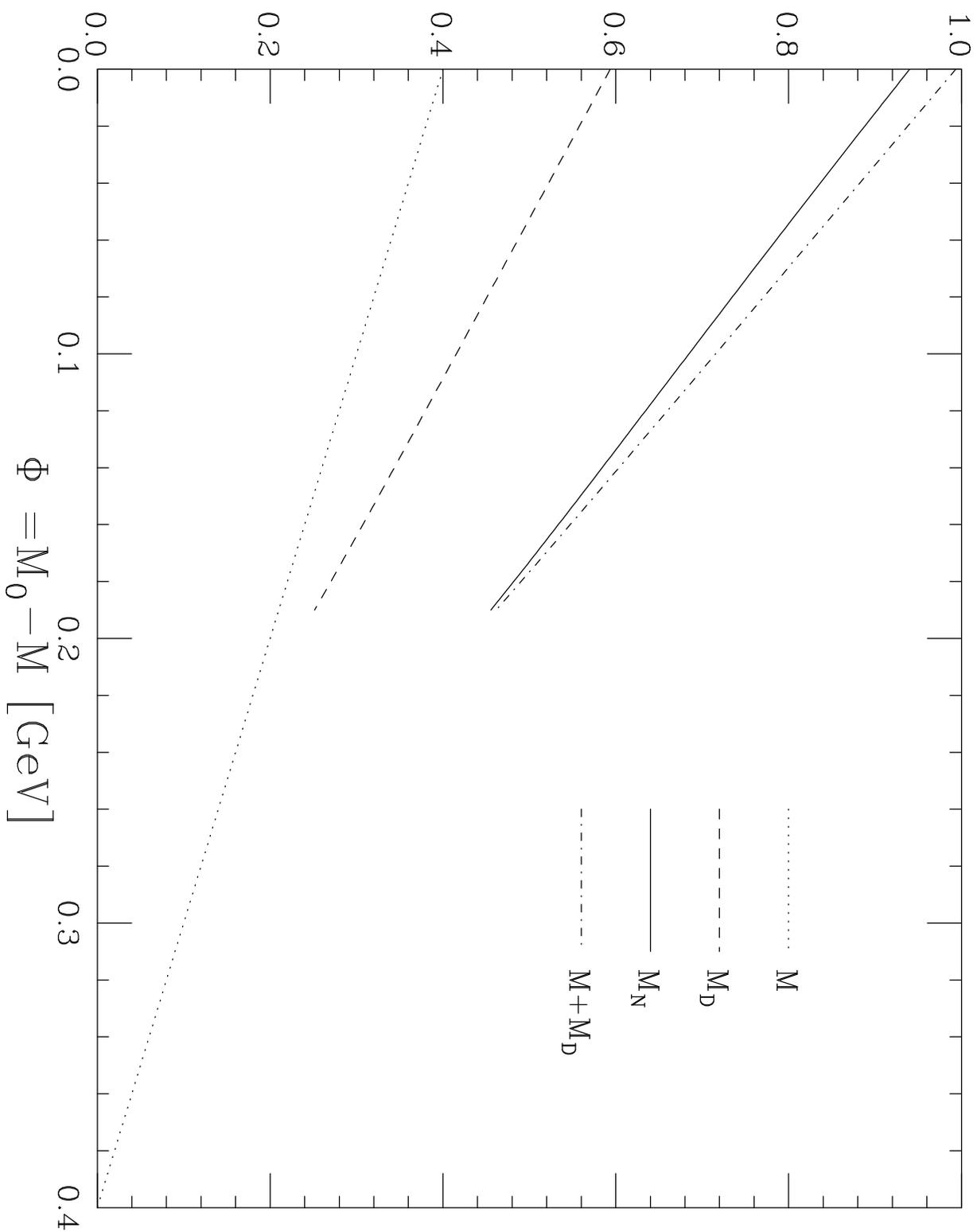
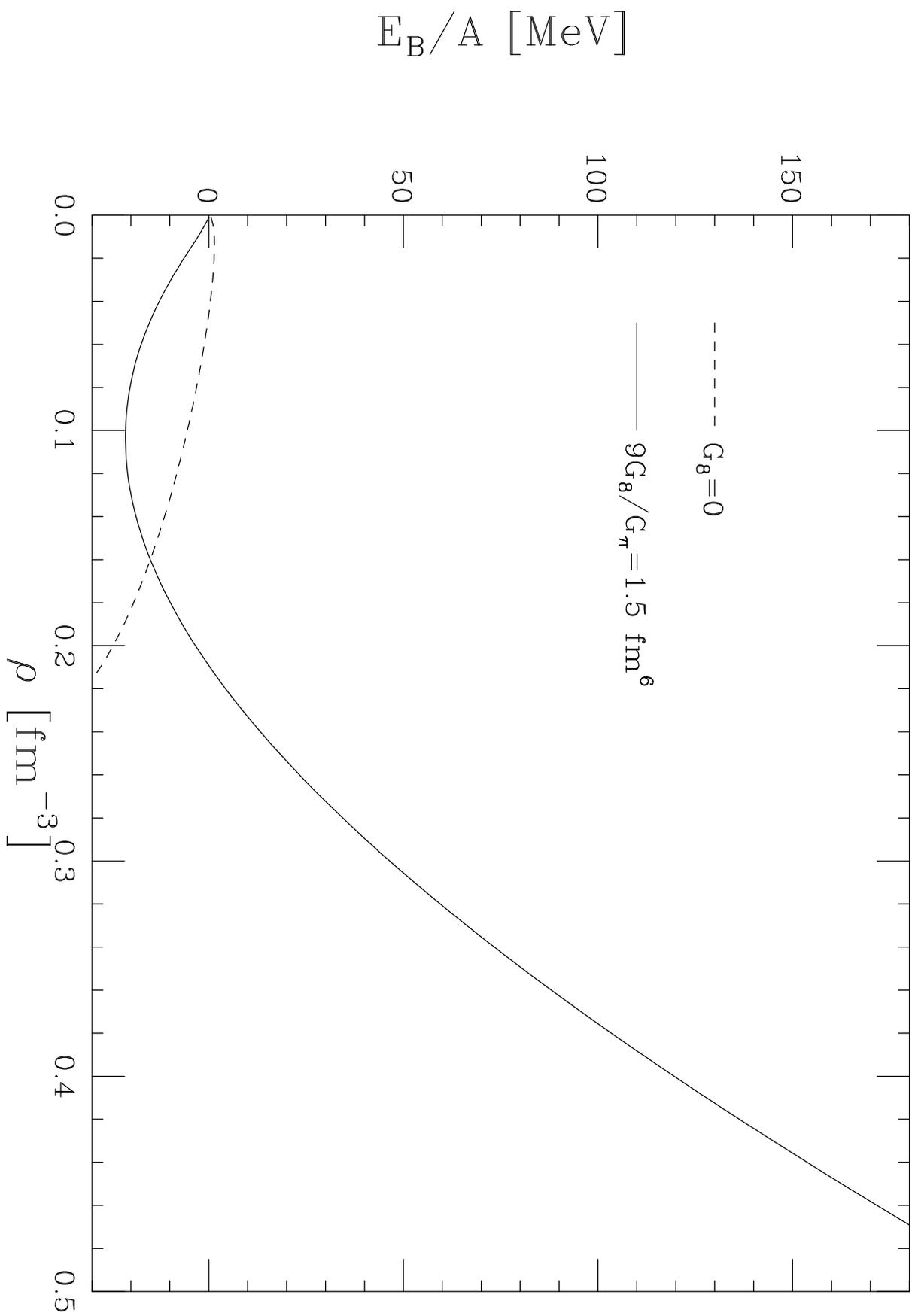
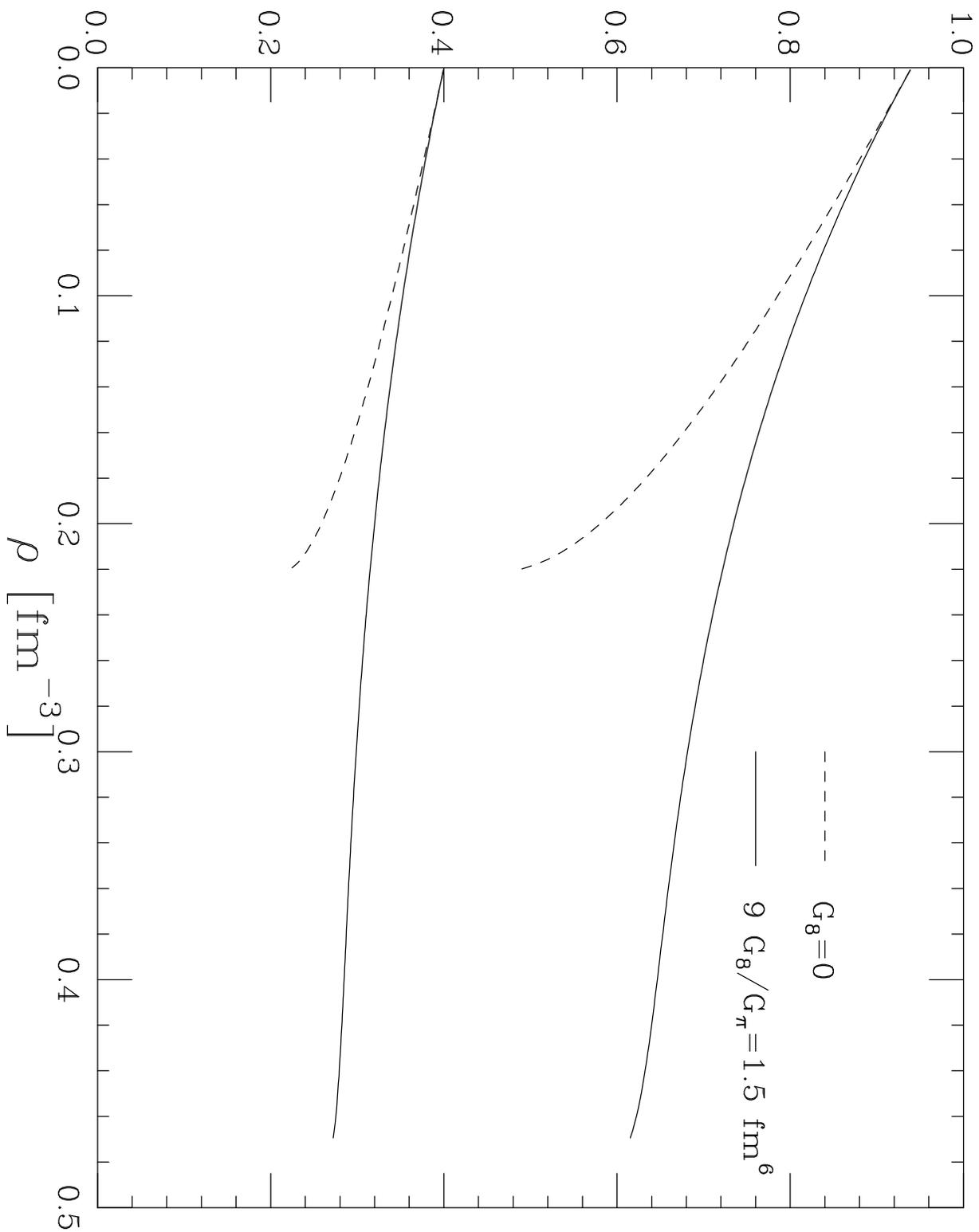
\end{document}